\documentclass[aps,prx,twocolumn,footnoteinbib,superscriptaddress]{revtex4-2}

\usepackage{braket}
\usepackage{ytableau}

\usepackage{url}
\usepackage{hyperref}
\usepackage{graphicx}
\usepackage{amsthm,amsfonts}
\usepackage{mathtools}
\usepackage{float}
\usepackage{soul}
\usepackage{amssymb}
\usepackage{dsfont}

\usepackage{tikz-cd}
\usepackage{quantikz}
\usepackage{adjustbox}
\usepackage[percent]{overpic} 

\newcommand{\rvline}{\hspace*{-\arraycolsep}\vline\hspace*{-\arraycolsep}}

\newcommand{\Id}{\mathds{1}}
\newcommand{\SU}{\mathrm{SU}}
\newcommand{\Wg}{\mathcal{W}g}
\newcommand{\Tr}{\mathrm{Tr}}
\newcommand{\Sym}{\mathfrak{S}}

\newcommand\numberthis{\addtocounter{equation}{1}\tag{\theequation}}

\newtheorem{definition}{Definition}
\newtheorem{lemma}{Lemma}
\newtheorem{theorem}{Theorem}
\newtheorem{corollary}{Corollary}
\newtheorem{example}{Example}

\DeclareSymbolFont{usualmathcal}{OMS}{cmsy}{m}{n}
\DeclareSymbolFontAlphabet{\mathcal}{usualmathcal}

\begin{document}

\title{All you need is spin: SU(2) equivariant variational quantum circuits based on spin networks}

\author{Richard D. P. East}
\email{rdp.east@gmail.com}
\affiliation{Xanadu, Toronto, ON, M5G 2C8, Canada}
\author{Guillermo Alonso-Linaje}
\affiliation{Xanadu, Toronto, ON, M5G 2C8, Canada}
\author{Chae-Yeun Park}
\email{chae.yeun.park@gmail.com}
\affiliation{Xanadu, Toronto, ON, M5G 2C8, Canada}
\affiliation{School of Integrated Technology, Yonsei University, Incheon 21983, Republic of Korea}
\affiliation{Department of Quantum Information, Yonsei University, Incheon 21983, Republic of Korea}
\affiliation{BK21 Graduate Program in Intelligent Semiconductor Technology, Incheon 21983, Republic of Korea}

\date{\today}

\begin{abstract}
Variational algorithms require architectures that naturally constrain the optimization space to run efficiently.
Geometric quantum machine learning achieves this goal by encoding group structure into parameterized quantum circuits to include the symmetries of a problem as an inductive bias.
However, constructing such circuits is challenging as a concrete guiding principle has yet to emerge.
In this paper, we propose the use of \textit{spin networks}, a form of directed tensor network invariant under a group transformation, to devise SU(2) equivariant quantum circuit ans\"{a}tze --- circuits possessing spin-rotation symmetry.
By changing to the basis that block diagonalizes the SU(2) group action, these networks provide a natural building block for constructing parameterized equivariant quantum circuits.
We prove that our construction is mathematically equivalent to other known constructions, such as those based on twirling and generalized permutations, but more direct to implement on quantum hardware.
The efficacy of our constructed circuits is tested by solving the ground state problem of SU(2) symmetric Heisenberg models on the one-dimensional triangular lattice and the Kagome lattice.
Our results highlight that our equivariant circuits boost the performance of quantum variational algorithms, indicating broader applicability to other real-world problems.
\end{abstract}

\maketitle

\section{Introduction} \label{sec:introduction}

Variational algorithms are prominent across physics as well as computer science with particularly fruitful applications in machine learning, condensed matter physics, and quantum chemistry~\cite{peruzzo2014variational,blei2017variational,carleo2019machine,cerezo2021variational}. 
In such areas, a parameterized function, often called an ansatz, is used to model a probability distribution or a quantum state, and parameters are optimized by minimizing a cost function.
However, this simple principle does not work without properly chosen ans\"{a}tze when dealing with a huge parameter space~\cite{wolpert1997no}.

For this reason, researchers often incorporate an \emph{inductive bias} into their algorithms~\cite{goyal2022inductive}.
An inductive bias is a prior knowledge about the system under investigation that can be included in the algorithm to restrict our function classes. Thus, the parameterized function favors a better class of outputs for a given target problem.
In classical machine learning, for example, it is known that the great success of convolutional neural networks (CNNs) is based on the fact that they contain `layers', essentially parameterized maps, which encode the idea that the content of an image does not change when shifted.
Specifically, these convolutional layers are (approximately) translation equivariant: When one shifts the input state by $n$ pixels horizontally and $m$ pixels vertically, the output is also shifted in the same way~\cite{cohen2016group,kondor2018generalization}.
Geometric deep learning naturally extends this framework to arbitrary groups~\cite{bronstein2021geometric}, suggesting the use of group equivariant layers for learning data with symmetric properties.
Neural networks consisting of group equivariant layers have indeed reported better performance for classifying images~\cite{cohen2016group}, point clouds~\cite{qi2017pointnet}, and in the modeling of dynamical systems~\cite{satorras2021n}. More broadly, they have also been used in a general variational context for tasks such as identifying the ground state of molecules~\cite{pfau2020ab}.

Recently, the idea of geometric machine learning has been combined with quantum machine learning (QML).
Generally speaking, QML algorithms~\cite{schuld2015introduction} hope to find an advantage over classical algorithms in ML tasks by exploiting the quantum nature of Hilbert space using parameterized quantum circuits.
Despite its potential, however, the trainability and generalization performance of QML algorithms without tailored circuit ans\"{a}tze often scale poorly, limiting their usability for more than tens of qubits~\cite{cerezo2022challenges}.
Because of this, recent studies introduced geometric quantum machine learning (GQML) as a guiding principle for constructing a quantum circuit ansatz. The literature shows these symmetry-informed circuits have been successful in offering better trainability and generalization performance~\cite{bowles2023backpropagation,west2022reflection,zheng2023sncqa,larocca2022group,zheng2022super,skolik2023equivariant,meyer2023exploiting,zheng2023speeding,sauvage2024building,heredge2024permutation,schatzki2024theoretical,nguyen2024theory}.

Group equivariant circuits also provide significant advantages in solving the ground state problem of many-body Hamiltonians, i.e., for variational quantum eigensolvers (VQEs)~\cite{peruzzo2014variational}.
The VQEs estimate the ground state energy of a given Hamiltonian by finding parameters of quantum circuits that minimize the Hamiltonian expectation value.
Still, one needs to choose a proper circuit ansatz to find the ground state accurately.
One of the design principles is letting the circuit have the same set of symmetry as the target Hamiltonian~\cite{wecker2015progress,hadfield2019quantum,wiersema2020exploring}.
As in the GQML, such a choice significantly improves the trainability and the accuracy of the algorithm~\footnote{Still, there is a caveat that such an advantage often comes with the expense of reduced expressivity. See, e.g., Ref.~\cite{park2024efficient} and discussion in Ref.~\cite{meyer2023exploiting}.}.
Among many group symmetries, the $U(1)$ symmetry is arguably the most studied in this setting as it is relevant to the particle number preservation law, which is essential in quantum chemistry~\cite{barkoutsos2018quantum,dallaire2019low,mcardle2020quantum,gard2020efficient,lee2018generalized}.

In both GQML and the VQEs, the $\SU(2)$ equivariant quantum circuits are far less studied~\cite{ragone2022representation,nguyen2024theory,meyer2023exploiting,zheng2023speeding}.
The symmetry group $\SU(2)$ is particularly interesting as it naturally arises in quantum systems with rotational symmetry.
For example, the Heisenberg XXX model defined on a lattice and the prominent Affleck–Kennedy–Lieb–Tasaki (AKLT) model~\cite{affleck2004rigorous} have the $\SU(2)$ symmetry.
It is also essential in QML for some data classes, such as point clouds, where the underlying properties of the data (e.g., whether the given input is a sphere or donut) do not depend on rotation.

However, existing literature does not provide a constructive method for implementing the $\SU(2)$ equivariant quantum circuit ans\"{a}tze.
For example, Ref.~\cite{meyer2023exploiting} proposed twirling as a method for obtaining generators of equivariant gates,
but computing this twirling formula for a many-qubit gate is highly non-trivial as it involves the summation over the symmetric group (thus over $n!$ terms).
In contrast, Ref.~\cite{zheng2023speeding} showed that a certain form of elements in an algebra generated by the symmetric group (formally written as $\mathbb{C}[\Sym_n]$ where $\Sym_n$ is a symmetric group) can be seen as $\SU(2)$ equivariant quantum circuits.
Nonetheless, these circuits do not admit a simple decomposition into few-qubit gates (implementable on quantum hardware).
Those constructions do not clearly show the set of linearly independent generators of the equivariant gates, either.

In this paper, we propose an alternative approach to construct $\SU(2)$ equivariant circuits. Our circuit ans\"{a}tze, dubbed \emph{spin-network circuits}, are inspired by spin networks, $\SU(2)$ equivariant tensor networks~\cite{penrose1971angular}. 
A core tool for us will be the \emph{Schur} gate (or map; we will use these terms interchangeably) that sends us from a qubit basis to a spin-basis. 
For example, it provides the following mapping for two qubits: $\ket{J=0,J_z=0} = \ket{01}-\ket{10}$, $\ket{J=1,J_z=1} = \ket{00}$, $\ket{J=1,J_z=0} = \ket{01}+\ket{10}$, and $\ket{J=1,J_z=-1} = \ket{11}$ where $J$ is the total angular momentum of two qubits and the $J_z$ is its $z$-direction component.
The advantage of this basis is that it leaves the matrix representations block-diagonal with respect to the total angular momenta \cite{kirby-practical-schur}.
We then further utilize this basis to parameterize all $\SU(2)$ equivariant maps, or equivalently, the commutant of $\SU(2)$ actions (see Sec.~\ref{sec:rep_equivariant} for details).

In addition, we prove that our circuit is mathematically equivalent to other constructions using the representation theory of $\SU(2)$.
In particular, we prove that both our gates and gates from the twirling formula~\cite{nguyen2024theory,meyer2023exploiting} can be written in the form of generalized permutations as introduced in Refs.~\cite{zheng2022super,zheng2023speeding}.
When restricted to unitary operators, all three constructions give the same set of gates.
Our main theoretical tool is the Schur-Weyl duality, which roughly posits a duality between $\SU(2)$ and the symmetric group $\Sym_n$.

We additionally show that the proposed three-qubit gates can be useful for solving a real-world problem with supporting numerical results for $\SU(2)$ symmetric models.
We choose the problem of finding the ground state of $\SU(2)$ symmetric Hamiltonians as it provides a better benchmark platform for classically simulated QML models (with $\sim 20$ qubits).
In particular, we solve the Heisenberg XXX model on one-dimensional triangular and Kagome lattices, which have the $\SU(2)$ symmetry but are tricky for Monte Carlo-based classical algorithms due to the sign problem~\cite{klassen2019two,hangleiter2020easing}.
We show that our circuit ans\"{a}tze give accurate ground states with a common parameter optimization technique which demonstrates the efficiency of our method and justifies the use of our $\SU(2)$ equivariant circuits for appropriately symmetric variational, QML problems, and quantum MaxCut~\cite{gharibian2019almost} more generally.

The paper is organized as follows. In Sec.~\ref{sec:preliminaries}, we introduce the preliminaries needed to understand the other sections: The representation theory for $\SU(2)$, spin coupling, and spin networks. In Sec.~\ref{sec:spin-network-circuits}, we introduce our ans\"{a}tze termed \emph{spin-network circuits}, which are parameterizable unitary quantum circuits that are also spin networks.
To this end, the Schur gate will be introduced, a core technical component in creating our parameterizations.
We also concretely present the two and three-qubit unitary \emph{vertex} gates.
In Sec.~\ref{sec:rep_equivariant}, we show that all $\SU(2)$ equivariant unitaries are a form of generalized permutation. 
This directly connects the work here with that on permutational quantum computing (PQC)~\cite{jordan2009permutational,havlivcek2018quantum} and in particular PQC+ as outlined in Ref.~\cite{zheng2023speeding}.
We also discuss the relation with the twirling method introduced in Ref.~\cite{meyer2023exploiting} showing how all $\SU(2)$ equivariant gates, i.e., generalized permutations, are the same as the set of all unitary gates generated by twirled Hermitian operators.
Next, in Sec.~\ref{sec:simulations}, we present the efficacy of the introduced vertex gates by solving the Heisenberg XXX model defined on one-dimensional triangular and two-dimensional Kagome lattices.
We then discuss the implications of our results and the connections to PQC and PQC+ in Sec.~\ref{sec:connections} and conclude with a short remark in Sec.~\ref{sec:conclusion}.

\section{Preliminaries}\label{sec:preliminaries}
\paragraph{Groups and their representations}
We are interested in equivariant quantum gates under the $\mathrm{SU}(2)$ group transformation. 
The group $\mathrm{SU}(2)$ itself is part of a larger class of groups known as $\SU(d)$ which is a set of $d\times d$ unitary matrices with determinant 1. 

Throughout the paper, we use the following terminologies.
First, an $\SU(2)$-\textit{equivariant} gate acting on $k$ qubits is a gate $T$ satisfying
\begin{align}
    T U^{\otimes k} = U^{\otimes k} T
\end{align}
for all $U \in \SU(2)$. Likewise, the whole circuit $C$ acting on $n$ qubits is $\SU(2)$-equivariant if $C U^{\otimes n} = U^{\otimes n} C$.
Moreover, a quantum state $\ket{\psi}$ is $\SU(2)$-\textit{invariant} if
\begin{align}
    U^{\otimes n} \ket{\psi} = \ket{\psi}.
\end{align}
An operator $O$ is $\SU(2)$-invariant (or symmetric) if 
\begin{align}
    (U^\dagger)^{\otimes n}OU^{\otimes n} = O,
\end{align}
for all $U\in \SU(2)$. The Heisenberg interaction $X\otimes X+Y\otimes Y + Z \otimes Z$ is a classical example, where $\{X,Y,Z\}$ are Pauli operators.
One can easily find that a circuit composed of $\SU(2)$-equivariant local gates (i.e., $k=O(1)$) is $\SU(2)$ equivariant, but the converse may not be true~\cite{marvian2022restrictions}.

Using an $\SU(2)$ equivariant circuit $C$, satisfying $CU^{\otimes n} = U^{\otimes n}C$, one can create an $\SU(2)$-invariant output state given an $\SU(2)$-invariant input state. 
If $|\psi_0 \rangle$ is an input state satisfying $|\psi_0 \rangle = U^{\otimes n}|\psi_0 \rangle$ (we will see an example of such a state in Sec.~\ref{sec:spin-network-circuits}), we have
\begin{align}
    U^{\otimes n} C \ket{\psi_0} = C U^{\otimes n} \ket{\psi_0} = C \ket{\psi_0}.
\end{align}
Thus, such a circuit $C$ can be used for learning tasks involving rotationally invariant data, e.g., finding ground states of $\SU(2)$-invariant Heisenberg (XXX) spin models or classifying point sets.

The symmetry we consider here is tightly connected to groups and their representations.
Recall that a group $G$ is a set with a map acting on two of its elements $\cdot: G \times G \rightarrow G$ such that there is an identity $e \in G$ such that $g \cdot e = e\cdot g = g$ for all $g\in G$, the operations are associative $g_1\cdot (g_2 \cdot g_3) = (g_1\cdot g_2) \cdot g_3$ for all $g_1,g_2,g_3 \in G$, and for any $g \in G$, an inverse element $g^{-1} \in G$ exists such that $g\cdot g^{-1} = e$.
It is also natural to consider the action of a group on a vector.
For example, a rotation $\mathrm{Rot} \in \mathrm{SO}(3)$ acts on a three-dimensional (real) vector and transforms it.
This type of action (on a vector space) is called a \textit{representation} of a group.

Formally speaking, a group representation is a map $R:G \rightarrow \mathrm{GL}(V)$ from the group to the space of invertible linear maps on a vector space $V$ (or equivalently, invertible matrices of dimension $N$ if $\dim(V)=N$) such that $R(g_1\cdot g_2) = R(g_1)\cdot R(g_2)$. In essence, it is a map from the group to linear maps that preserves the group structure. 
For a system with a single qubit, a simple map $R(U)=U$ for $U \in \SU(2)$ already defines a representation.
One can readily extend this representation to an $n$-qubit system by defining $\widetilde{R}(U) = U^{\otimes n}$, which is also a representation (as $\widetilde{R}(U_1 U_2) = (U_1 U_2)^{\otimes n} = U_1^{\otimes n} U_2^{\otimes n} = \widetilde{R}(U_1)\widetilde{R}(U_2)$).
We can then see that to find $\SU(2)$ equivariant gates for an $n$-qubit system, we must pay attention to the representation $\widetilde{R}$.

Studying the representation of symmetry introduces the concept of \textit{irreducible representations} (irreps, for short). Firstly, a sub-representation $R|_W$ of $R$ is a representation on $W$ which satisfies $R(g)W = \{R(g)w: w \in W\} \subseteq  W$ for all $g \in G$, where $W$ is a subspace of $V$. Then we say a representation $R:G\rightarrow \mathrm{GL}(V)$ is irreducible if it does not have any non-trivial sub-representations, i.e., if $W \leq V$ and $R(g)W = \{R(g)w: w \in W\} \subseteq  W$ for all $g \in G$, then $W = 0$ or $W = V$. Thus, we may find a structure of equivariant gates by decomposing an $n$-qubit system to irreducible representations (which is always possible by the Peter–Weyl theorem).
As we shall see, the \textit{Schur map} sends equivariant operators into a block-diagonal form. This form will allow us to explicitly design such maps.

\paragraph{From qubits to spins}

A spin is an irreducible representation of the $\SU(2)$ group.
This vector space is spanned by basis vectors  $\{\ket{J, J_z}: -J \leq J_z \leq J\}$ where $2J$ is an integer (e.g., $J=0$, $J=\frac{1}{2}$, $J=1$, $J=3/2$, etc.).
Physically, $J$ and $J_z$ correspond to the quantized total angular momentum and the angular momentum in the $z$-direction, respectively (though the $z$-direction is a conventional choice, any would do).
For each allowed value of $J$, we call the corresponding vector space a spin-$J$ system.

\begin{figure}[t]
    \centering
	\includegraphics[width=0.92\linewidth]{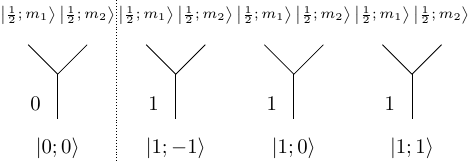}
    \caption{
    Graphical presentation of the basis constructed by combining angular momentum of two spin-$\frac{1}{2}$ systems and the possible outcomes of total and $z$-directed angular momenta.
    These can be seen as two spin networks, each consists of a single tensor, corresponding to the two possible total angular momentum values on the bottom edge, with specific $\ket{J;J_z}$ states chosen for the bottom edges' Hilbert spaces.
    Note that we just expanded one spin network (tensor) for the total outcome angular momentum $J=1$ to three graphs, where each output label corresponds to the tensor index of the spin network. For example, the graph for $\ket{1;0}$ represents the Clebsch-Gordan coefficient $c^{1,0}_{1/2,m_1,1/2,m_2}$.
    }
    \label{fig:2-couple}
\end{figure}

A qubit is naturally identified as a spin-$\frac{1}{2}$ particle, by a mapping $\ket{0}=\ket{J=\frac{1}{2};J_z=\frac{1}{2}}$ and $\ket{1}=\ket{J=\frac{1}{2};J_z=-\frac{1}{2}}$.
When we take two qubits, we are thinking of the basis elements $\{\ket{00},\ket{01},\ket{10},\ket{11}\}$.
Consider the angular momentum of two qubits (or two spin-$\frac{1}{2}$ particles, equivalently).
It is well known that when one considers two spin-systems of momenta $J_1$ and $J_2$ in terms of their joint angular momentum, the possible total angular momentum $J$ measurements range from  $J=|J_1-J_2|$ to $J_1+J_2$.
Thus, two qubits have the two total angular momentum possibilities of $J=0$ and $J=1$. To get the full basis, we must include the possible $J_z$ values ranging from $-J$ to $J$ in steps of 1~\cite{hall2013lie}.
In general, we can always move from a basis of qubits to a basis of angular momenta by considering the pairwise coupling of qubits and subsequent spins, which amounts to considering the possible angular momentum outcomes of a measurement of each pairing.
This coupling scheme is depicted in Fig.~\ref{fig:2-couple}.

\begin{figure}[t]
    \centering
    \includegraphics[width=0.95\linewidth]{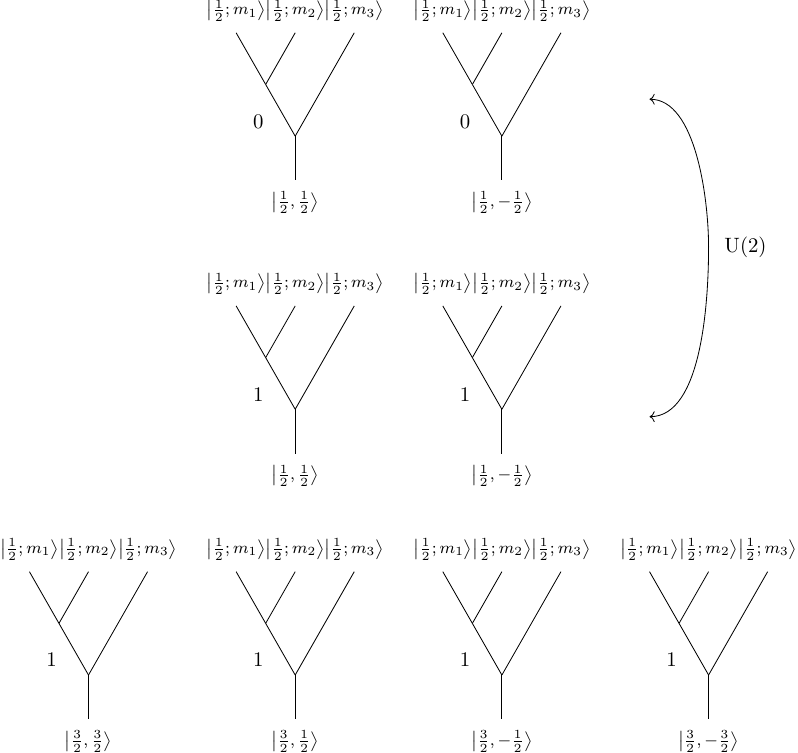}
    \caption{Graphical depiction of a coupling basis of three qubits, where the pairwise coupling of the spaces proceeds from the left (other possibilities give alternative bases). Each row of trees is indexed by the possible total angular momenta that can occur for each composition of two systems. The elements in the rows correspond to the different states, giving a final $J_z$ value on the spaces at the bottom of the trees. Note how the top two rows of diagrams index spaces with the same total angular momentum at the base but that the patterns of coupling that form them are distinct. In Sec.~\ref{sec:rep_equivariant}, we will see that this allows for the mixing of such states because $\SU(2)$ equivariant maps cannot distinguish the two spin coupling structures. Note that in the absence of specifying the $J_z$ values, the set of diagrams on each row correspond to three separate spin networks as the $\SU(2)$ invariance on three-valent networks reduces to spin-coupling rules; this is discussed in more detail in Appendix~\ref{sect:q-geom}.}
    \label{fig:trees}
\end{figure}

For more than two spins, we will have a choice of the order in which we do this. The different orders of pairing the spin systems amount to different bases (as they correspond to different choices of complete measurements), which we can describe by branching tree-like structures. In Fig.~\ref{fig:trees}, we can see this for three qubits.

In later discussions, we will use $\mathfrak{J}_{J} = \mathbb{C}^{2J+1}$ to denote a spin-$J$ system. For example, $\mathfrak{J}_{1/2}=\mathbb{C}^2$ is a vector space for spin-$\frac{1}{2}$ system, i.e., a qubit.

\paragraph{Spin networks}
We now consider a generalization of equivariant gates using multi-linear maps.
Let us first recall properties of spin-$1/2$ kets and bras under $g \in \SU(2)$:
\begin{align}
    \ket{a} &\xrightarrow{g} g \ket{a} \\
    \bra{b} &\xrightarrow{g} \bra{b} g^\dagger,
\end{align}
where $g = e^{-i \phi \pmb{\sigma} \cdot \hat{n}/2} \in \SU(2)$. Here, $\pmb{\sigma} = \{\sigma_x, \sigma_y, \sigma_z\}$ is a vector of $2\times 2$ Pauli matrices, $\hat{n}$ is a normal vector indicating the direction of the rotation, and $\phi$ is the rotation angle. 

By identifying kets as vectors and bras as dual vectors, we can generalize the above principle by considering an arbitrary spin-$J$ system given as $V = \mathfrak{J}_{J} =\mathbb{C}^{2J+1}$.
Then $\ket{a} \in V$ and $\bra{b} \in V^*$ changes to
\begin{align}
    \ket{a} &\xrightarrow{g} R(g) \ket{a} \\
    \bra{b} &\xrightarrow{g} \bra{b} R(g)^\dagger
\end{align}
under the group transformation, where $R(g)$ is a representation of $g \in \SU(2)$.
Specifically, it is a $2J+1$ by $2J+1$ unitary matrix given by $e^{-i \phi \pmb{J} \cdot \hat{n}}$ which is a representation of $e^{-i \phi \pmb{\sigma} \cdot \hat{n}/2} = g \in \SU(2)$.
Here, $\pmb{J}=\{J_x, J_y, J_z\}$ is a vector of $2J+1$ by $2{J}+1$ spin matrices satisfying $[J_a, J_b] = i \epsilon_{abc} J_c$ for all $a, b, c \in \{x,y,z\}$ where $\epsilon_{abc}$ is the Levi-Civita symbol.

The above principle also induces group transformation formulas for other expressions.
For example, one can see that the inner product $\langle a|b \rangle$ is invariant under the group transform as
\begin{align}
    \langle b | a \rangle \xrightarrow{g} \bra{b} R(g)^\dagger R(g) \ket{a} = \langle b | a \rangle.
\end{align}
Note that the last equality is obtained as $R(g)$ is unitary.
Next, let us consider a linear map $T: V \rightarrow V$. 
As $T$ can be written as $T = \sum_{ij} t_{ij} \ket{i} \bra{j} \in V \otimes V^*$, we know it changes to
\begin{align}
    T \xrightarrow{g} R(g) T R(g)^\dagger
\end{align}
under the transformation.

We now add a constraint that a linear map $T$ also preserves the group structure. In other words, we require $T$ to satisfy
\begin{align}
    R(g) (T\ket{a}) = T (R(g) \ket{a})
\end{align}
for all $g \in G$ and $\ket{a} \in V$, which implies that $R(g)^\dagger T R(g) = T$ (or equivalently, $ T  = R(g)T R(g)^\dagger$).
As $R(g) T R(g)^\dagger$ is nothing but $T$ after the group transformation, a linear map preserving the group structure is a matrix that is invariant under the group transformation (given by conjugation with $R(g)$).

\begin{figure}[t]
    \centering
    \includegraphics[width = 0.85 \linewidth]{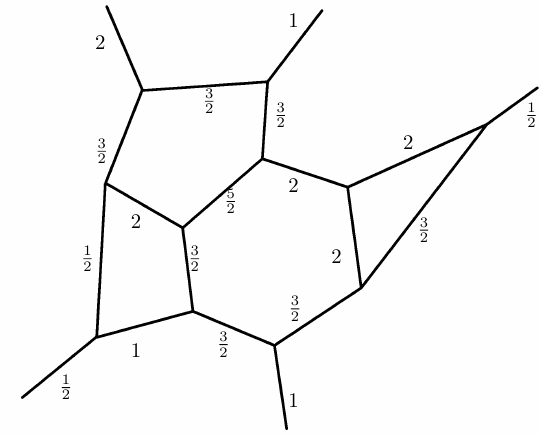} 
    \caption{A three-valent spin network typically presented in the broader literature: an edge-labeled graph (though directed, this is often suppressed in depictions since the spaces are isomorphic). In the three-valent case, the edge labels are spins such that around any vertex they meet the Clebsch-Gordan conditions $j_1+j_2+j_3\in\mathbb{N}$ and $ \lvert j_1-j_2\rvert\leqslant j_3\leqslant j_1+j_2.$ which can be shown to exactly match when the vertex is an invariant subspace of $\SU(2)$ (See Appendix~\ref{sect:q-geom} for more details).}
    \label{fig:3-valent-spin-network}
\end{figure}

One may further extend this property to multilinear maps (tensors). For example, a two-qubit gate is a linear map $T$ between $V^{\otimes 2}$ and $V^{\otimes 2}$ (where $V=\mathfrak{J}_{1/2}=\mathbb{C}^2$ in the standard formulation).
If we add the equivariant condition to this gate, i.e., $R(g)^{\otimes 2} T = T R(g)^{\otimes 2}$, this is nothing but the condition for a group-structure preserving map.
As a two-qubit gate $T$ can be considered as an element of $V^{\otimes 2} \otimes (V^*)^{\otimes 2}$, $T$ becomes
\begin{align}
    T \xrightarrow{g} R(g)^{\otimes 2} T (R(g)^\dagger)^{\otimes 2} = T,
\end{align}
under the group transformation, where the last equality is from the equivariant condition. Formally, the space of tensors in $V^{\otimes n} \otimes (V^*)^{\otimes m}$ that preserve the symmetry is often written as $\mathrm{Inv}_{\rm SU(2)}(V^{\otimes n} \otimes (V^*)^{\otimes m})$.
Thus, a one-to-one correspondence exists between group-structure preserving maps and group-invariant tensors.
In other words, if we consider a general (possibly non-unitary) linear map between $V^{\otimes n}$ and $V^{\otimes m}$ (where $n$ and $m$ can be different integers), preserving the group structure,
it can be seen as a group-invariant tensor with $n$ input legs and $m$ output legs~\cite{singh2010tensor,singh2012tensor} (often called a tensor of type $(n,m)$).

Now, we consider a tensor network that consists of $\SU(2)$ invariant tensors with contraction edges that run over irreps of $\SU(2)$.
This special type of network is called a ``spin network''; an example from the broader literature can be seen in Fig.~\ref{fig:3-valent-spin-network}. These were originally introduced by Penrose~\cite{penrose1971angular} in the very different context of a combinatorial derivation of space-time.
In modern physics, they are typically discussed as the basis of quantized space in the covariant formulation of loop quantum gravity~\cite{rovelli1995spin}.
Roughly, a spin network is a directed graph where each edge has an associated spin, and each vertex $v$ has an associated equivariant map from the tensor product of the incoming spins to the tensor product of the outgoing spins.
Formally, we describe this as a graph detailing the connectivity of vertices $v$ with incoming edges $e_{in}$ and outgoing ones $e_{out}$ such that for every vertex, there is an associated map $T_v$ such that $T_v \in \bigotimes_{i\in e_{in}}\bigotimes_{o\in e_{out}} \mathfrak{J}_{j_{i}} \otimes \mathfrak{J}_{j_{o}}^*$,  where $\mathfrak{J}_{j_{i}}$ and $\mathfrak{J}_{j_{o}}$ are the incoming and outgoing respective Hilbert spaces.
We further require $T_v$ to satisfy the equivariant condition
\begin{align}
&\bigotimes_{i\in e_{in}} \bigotimes_{o\in e_{out}} T_v \left(R_{j_{i}}(g)\mathfrak{J}_{j_{i}} \otimes  \mathfrak{J}_{j_{o}}^*\right)\nonumber \\
&= \bigotimes_{i\in e_{in}} \bigotimes_{o\in e_{out}} T_v \left( \mathfrak{J}_{j_{i}} \otimes  R_{j_{o}}(g)^\dagger \mathfrak{J}_{j_{o}}^* \right), \, \forall g\in G \text{ and } \forall v, \label{eq:vertex}
\end{align}
where $R_{j_i}(g)$ and $R_{j_o}(g)$ are the representations of the group element $g$ acting on the $\mathfrak{J}_{j_{i}}$ and $\mathfrak{J}_{j_{o}}$, respectively.
From the discussion above, each map associated with a vertex, $T_v$, can be regarded as a group-invariant tensor.
In this way, spin networks are tensor networks where the composing tensors are elements in the invariant subspaces of a group, and the contraction is over spin spaces of size $2J+1$.
For a more detailed description of these objects, we direct the reader to Appendix \ref{sect:q-geom}.
For our interests, it is sufficient to say that we can build a quantum circuit that is inherently $\SU(2)$ equivariant by restricting to specific spin networks whose vertices can be interpreted as parameterized qubit unitaries.

Within the literature, spin networks that form binary trees have been particularly prominent.
The simplest example is seen in Fig.~\ref{fig:2-couple}, where we ignore the specification of the $J_z$ state at the bottom and focus only on the total angular momentum (so there are just two unique diagrams from this perspective).
A more general example is provided by Fig.~\ref{fig:trees}, where we have three spin spaces coming together, which naturally leads to three possible spin networks, specifically one for each row.
The columns are not different networks because they amount to fixing a choice of $J_z$ value on one edge, which is a choice of contraction index (i.e., final projection).
Thus, such a fixing does not alter the spin spaces in the network definition.
It should be noted that spin networks have previously been considered in the broader quantum information literature as diagrammatic qubit maps and as variational maps for numerical investigations of lattice quantum gravity on quantum computers Refs.~\cite{east2021spin, mielczarek2019spin,czelusta2021quantum,czelusta2023quantum}, though never as general $\SU(2)$ equivariant variational ans\"{a}tze.

\section{Spin-network circuits}\label{sec:spin-network-circuits}

In this section, we outline circuit ans\"{a}tze designed based on the principles of spin networks whose utility will be shown further below in Sec.~\ref{sec:simulations} using concrete numerical simulations.
Our circuits, termed \emph{spin-network circuits}, are a specific form of spin network, so they are explicitly $\SU(2)$ equivariant.
They are spin networks where all vertices have an even number of external wires, and every wire in the network is spin-$\frac{1}{2}$, and so are formed of qubits.
Among all external wires for each vertex, half are inputs, and the other half are outputs; the combination of these vertices amounts to a quantum circuit.
For this reason, when viewed as a quantum circuit, we refer to the vertices as \emph{vertex gates}.
Critically, the vertices of a spin network are equivariant maps between the input and output edges, which is a direct consequence of the definition given in Eq.~\eqref{eq:vertex}. This means the resultant circuit is also equivariant.
An important property of spin networks with vertices with more than three edges is that they can be parameterized (see Appendix \ref{sect:q-geom}). By training over these parameters, we thus arrive at a trainable equivariant network.

\paragraph{Schur gate and two-qubit vertex gate}

The simplest spin-network circuit is built from vertex gates acting solely on two qubits. To understand the structure of this gate and its later generalizations, we first require the two-qubit Schur gate as a prerequisite~\cite{havlivcek2019classical}:
\begin{align}
    S_2 = \begin{pmatrix}
    1 & 0 & 0 & 0\\
    0 & \frac{1}{\sqrt{2}} & \frac{1}{\sqrt{2}} & 0\\
    0 & 0 & 0 & 1\\
    0 & \frac{1}{\sqrt{2}} & -\frac{1}{\sqrt{2}} & 0
    \end{pmatrix} \label{eq:schur_two}
\end{align}

This gate is a unitary operator that maps the computational basis of two qubits to the spin basis of their combined $J$ and $J_z$ angular momenta.
As qubits can be seen as spin-$\frac{1}{2}$ spaces, with spin-up and spin-down being assigned to $0$ and $1$ respectively, then qubit registers correspond to tensor products of spin-$\frac{1}{2}$ irreps.
While these are individually irreducible, their tensor product is not and can be block-diagonalized into irreducible components. 
In the case of two qubits, it is often typical to write that $\mathfrak{J}_{\frac{1}{2}} \otimes \mathfrak{J}_{\frac{1}{2}}\simeq \mathfrak{J}_{0} \oplus \mathfrak{J}_{1}$ which says that a tensor product of two spin-$\frac{1}{2}$ spaces is isomorphic to the direct sum of a spin-$0$ and a spin-$1$ space telling us that there is a unitary map between them. The two-qubit Schur gate performs exactly this map. Looking at this in terms of the computational basis, the two-qubit Schur gate maps the computational basis states to the following basis (where we often drop the normalization in later exposition):  $\ket{J=1,J_z=1} = \ket{00}$, $\ket{J=1,J_z=0} = \frac{1}{\sqrt{2}}(\ket{01}+\ket{10})$, $\ket{J=1,J_z=-1} = \ket{11}$, and $\ket{J=0,J_z=0} = \frac{1}{\sqrt{2}}(\ket{01}-\ket{10})$, which is occasionally referred to as the triplet/singlet basis.
In general, though trivially in the two-qubit case, we can say that the two-qubit Schur map sends us to the sequentially coupled basis of two qubits exactly as depicted in Fig.~\ref{fig:2-couple}. As was discussed in Sec.~\ref{sec:preliminaries} above, this amounts to two different binary spin networks with the $J_z$ values specified on the base~\cite{marzuoli2005computing}.

The two-qubit Schur gate from Eq.~\eqref{eq:schur_two} is the simplest Schur map that sends the tensor product of qubits to the direct sum of spins.
Precisely, the general form of the Schur map follows the prescription:
\begin{equation}\label{eq:schur-map}
   S_n: \mathfrak{J}_{\frac{1}{2}}^{\otimes n}\to \bigoplus_{k} \mathfrak{J}_k
\end{equation}
where we understand $\mathfrak{J}_{\frac{1}{2}}^{\otimes n}$ as the Hilbert space corresponding to $n$ qubits and $k$ ranges over the irreducible representations of $\SU(2)$ that make up the space in the spin-basis.
Note that \emph{irreps can repeat}, in which case we say there is a multiplicity.
In the next section, we will use the Schur-Weyl duality to compute the multiplicity of each irrep from the Schur map $S_n$ for arbitrary $n$.

The matrix elements of the Schur map can be obtained by using \emph{Clebsch-Gordan coefficients} and coupling paths of qubits. 
Each Clebsch-Gordan coefficient $\left\langle j_1 m_1 j_2 m_2 \mid J M\right\rangle=c_{j_1, m_1; j_2, m_2}^{J, M}$ corresponds to the projection of two particular spin-states into their combined angular momenta.
Thus, its matrix entries correspond to the Clebsch-Gordan coefficients that result from projecting coupled spin systems (specifically one spin-$\frac{1}{2}$ with whatever angular momentum that previous spin-couplings have reached) into a particular total $J$ value.
Each coefficient that gets multiplied corresponds to a vertex in the coupling diagrams that index each of the spin-basis elements (such as those seen in Fig.~\ref{fig:trees}), i.e., each element of the Schur map can be obtained by multiplying the Clebsch-Gordan coefficients associated with each vertex of the spin-coupling diagram.

As an example, let us consider the three-qubit case.
Here each element in the matrix of the Schur map corresponds to  $c^{j',m'}_{j_1,m_1;j_2,m_2}c^{J,M}_{j',m';j_3,m_3}$ for some choice of $j' \in \{0, 1\}$ and $-j' \leq m' \leq j'$.
Here $j'$ stands for the resulting spin from coupling the first two qubits, which leads to possible total spin momenta $j'=0$ and $j'=1$.
In the following, we focus on the spin-$0$ case ($j' = 0$).
This corresponds to the coefficient $c^{0,0}_{\frac{1}{2},m_1;\frac{1}{2},m_2}$. When we, in turn, couple with the third qubit the only possible outcome for the total angular momentum is $\frac{1}{2}$, so the combined coupling coefficient for these total angular momenta is $c^{0,0}_{\frac{1}{2},m_1;\frac{1}{2},m_2}c^{\frac{1}{2},m}_{\frac{1}{2},m_1;0,0}$.
These choices single out a particular recoupling path with associated final $J_z$ values on the root (as seen in Fig.~\ref{fig:2-couple}) and so a row in the matrix. The computational basis, equivalently the $J_z$ values for the individual qubits, fixes the columns (for more on this, see Ref.~\cite{wills2023generalized}). For practical implementations, it is important to note that the Schur gate can be implemented in polynomial time, and the literature already contains examples of specific methods to do this~\cite{wills2023generalized,bacon2006efficient}.

In the case of two qubits, there is only a single coefficient to consider in each element of the matrix, and so we have the following:
\begin{align}\label{eq:schur}
    S_{2} &= \begin{pmatrix}
c^{1,1}_{\frac{1}{2},\frac{1}{2};\frac{1}{2},\frac{1}{2}} & c^{1,1}_{\frac{1}{2},\frac{1}{2}; \frac{1}{2},-\frac{1}{2}}& c^{1,1}_{\frac{1}{2},-\frac{1}{2}; \frac{1}{2},\frac{1}{2}}& c^{1,1}_{\frac{1}{2},-\frac{1}{2};\frac{1}{2},-\frac{1}{2}}\\
c^{1,0}_{\frac{1}{2},\frac{1}{2};\frac{1}{2},\frac{1}{2}} & c^{1,0}_{\frac{1}{2},\frac{1}{2}; \frac{1}{2},-\frac{1}{2}}& c^{1,0}_{\frac{1}{2},-\frac{1}{2};\frac{1}{2},\frac{1}{2}}& c^{1,0}_{\frac{1}{2},-\frac{1}{2};\frac{1}{2},-\frac{1}{2}}\\
c^{1,-1}_{\frac{1}{2},\frac{1}{2};\frac{1}{2},\frac{1}{2}} & c^{1,-1}_{\frac{1}{2},\frac{1}{2};\frac{1}{2},-\frac{1}{2}}& c^{1,-1}_{\frac{1}{2},-\frac{1}{2};\frac{1}{2},\frac{1}{2}}& c^{1,-1}_{\frac{1}{2},-\frac{1}{2};\frac{1}{2},-\frac{1}{2}}\\
c^{0,0}_{\frac{1}{2},\frac{1}{2};\frac{1}{2},\frac{1}{2}} & c^{0,0}_{\frac{1}{2},\frac{1}{2};\frac{1}{2},-\frac{1}{2}}& c^{0,0}_{\frac{1}{2},-\frac{1}{2};\frac{1}{2},\frac{1}{2}}& c^{0,0}_{\frac{1}{2},-\frac{1}{2};\frac{1}{2},-\frac{1}{2}}
\end{pmatrix} \nonumber \\
&= \begin{pmatrix}
1 & 0 & 0 & 0\\
0 & \frac{1}{\sqrt{2}} & \frac{1}{\sqrt{2}} & 0\\
0 & 0 & 0 & 1\\
0 & \frac{1}{\sqrt{2}} & -\frac{1}{\sqrt{2}} & 0
\end{pmatrix}
\end{align}
which indeed matches the definition of the two-qubit Schur gate in Eq.~\eqref{eq:schur_two}.

Once we are in the spin basis, we can elegantly construct the two-qubit vertex gate by applying a phase solely on the spin-0, or singlet, element $\ket{J=0,J_z=0}$ (see Lemma~\ref{lem:schur} below). Intuitively, suppose a map is $\SU(2)$ equivariant so that you can isolate and apply group representations before or after the map. In that case, the different spin-irreps should not interact under the mapping and remain differentiated -- as matrices. This is why the map is block diagonal in the spin basis. For the two-qubit case, up to a global phase, this amounts to just a phase on one of the spaces:
\begin{equation}\label{eq:block-p}
P_2(\theta)= \begin{pmatrix}
    \text{ } & \text{ }     & \text{ }     & \rvline & 0 \\
    \text{ } & \Id_3 & \text{ }     & \rvline & 0 \\
    \text{ } & \text{ }     & \text{ }     & \rvline & 0 \\
    \hline
    0 & 0 & 0 & \rvline & e^{i\theta} 
\end{pmatrix}
\end{equation}
In terms of spin networks, which we recall are equivariant maps, the Schur gate is sending us to the two possible coupling options. Two qubits coupling to spin-$0$ or to spin-$1$. 
The parameterized gate $P_2(\theta)$ applies a phase on the spin-$0$ network.
In Sec.~\ref{sec:rep_equivariant}, this structure completely characterizes the possible unitary equivariant maps. 
To understand how this phase manages to isolate only one part of the spin space, we need to look again at representations. 
In the spin basis we obtain after the Schur gate, any group representation (up to row permutation depending on your exact basis choices and Schur gate, which can vary a little in the literature) is block diagonal. Each individual block is associated with a particular total angular momentum $J$ \emph{and} a way of arriving at it by sequentially coupling spin-$1/2$s as seen in Fig.~\ref{fig:trees}.
In this way, given a tensor product of $n$-spins, each block corresponds to one of the $2J+1$ dimensional spin spaces of its direct product decomposition as seen in Eq.~\eqref{eq:schur-map}. As we now know, for the case of two qubits, we either have spin-0 or spin-1, and so this block decomposition resembles the following:

\begin{figure*}[t]
    \centering
	\includegraphics[width=0.65\linewidth]{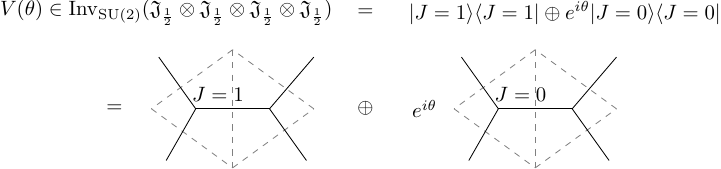}
    \caption{
		Depiction of a parameterized gate $V(\theta)\in \mathrm{Inv}_{\mathrm{SU(2)}}(\mathfrak{J}_\frac{1}{2}\otimes \mathfrak{J}_\frac{1}{2} \otimes \mathfrak{J}_\frac{1}{2} \otimes \mathfrak{J}_\frac{1}{2})$ in the basis that block diagonalizes $\SU(2)$ equivariant unitaries on two qubits and therefore is a four-valent spin-network vertex.
        It is composed of two three-valent spin networks indexed by the possible internal spin-$0$ or spin-$1$ edge. The solid lines represent the spin network with explicit intermediate $J$ values, and the dashed triangles separate the components of the map into their underlying 3-valent intertwiners (see Appendix \ref{sect:q-geom}).
	}
    \label{fig:param-basis}
\end{figure*}

\begin{equation}\label{eq:block}
\begin{pmatrix}
   & \rvline & 0 \\
  \text{spin-1} & \rvline & 0\\
   & \rvline & 0\\
\hline
0\text{ }\text{ }\text{ } 0  \text{ }\text{ }\text{ } 0 & \rvline & \text{spin-0}
\end{pmatrix}
\end{equation}
The block diagonal structure is critical for our $\SU(2)$ equivariant ans\"{a}tze. As we will see below, their general structure is to apply parameterized maps that act independently on blocks of different sizes (which are different irreducible representations) and as unitaries that mix those parts of repeated blocks of the same irreducible representation when they correspond to the same $J_z$ value. Indeed, this structure completely characterizes equivariant maps, as is shown below in Sec.~\ref{sec:rep_equivariant}.
As such, we can create an equivariant ansatz for $\SU(2)$, i.e., spin-rotation symmetry.

This leads us to the definition of a vertex gate.
\begin{definition}
The two-qubit vertex gate $V_2(\theta)$ is composed as follows:
\begin{equation}
\begin{quantikz}
& \gate[2]{V_2(\vec{\theta})} & \qw \\
& & \qw
\end{quantikz} =
\begin{quantikz}
& \gate[2]{S_2} & \gate[2]{P_2(\vec{\theta})} & \gate[2]{S_2^\dagger} & \qw \\
&               & & & \qw
\end{quantikz}
\end{equation}
where $S_2$ is the two qubit Schur gate and $P_{2}(\theta)$ is the controlled phase seen in Eq.~\eqref{eq:block-p}.
\end{definition}

\begin{figure}[b]
    \centering
	\includegraphics[width=0.85\linewidth]{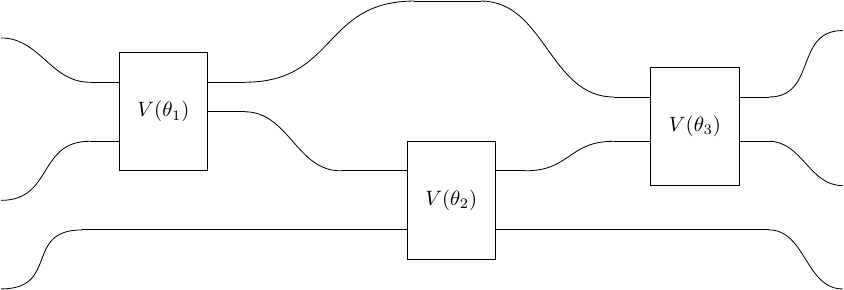}
    \caption{A four-valent spin-network circuit that can be trained over the free parameters in its vertex gates. The curved qubit wires highlight that such spin-network circuits are both spin networks and quantum circuits.}
    \label{fig:exp-2v}
\end{figure}

What we have created is specific two-qubit gates that live in the space of equivariant maps from, and to, the tensor product of two spin-$\frac{1}{2}$s; these can be seen depicted in Fig.~\ref{fig:param-basis}. 
These, by definition, are elements of the vertices of a four-valent spin network with edges fixed as qubits. We can see the spin network as corresponding to an operator formed by sequential gate operations as seen in Fig.~\ref{fig:exp-2v}

\paragraph{Three and more qubit vertex gates}

The three-qubit Schur gate, $S_3$,  transforms the tensor product of three spin-$1/2$ spaces to a direct sum of one spin-$1$ and two spin-$0$ spaces.
The matrix representation of $S_3$ can be obtained by computing the product of Clebsch-Gordan coefficients as described in Fig.~\ref{fig:trees}:

\begin{align}
    S_{3} &= (c^{j_4,m_4}_{j_1,m_1;j_2,m_2}c^{J,M}_{j_4,m_4;j_3,m_3}) \nonumber \\
    &= \begin{pmatrix} 1 & 0 & 0 & 0 & 0 & 0 & 0 & 0\\
                  0 & \frac{1}{\sqrt{3}} & \frac{1}{\sqrt{3}} & 0 & \frac{1}{\sqrt{3}}& 0 & 0 & 0 \\
                   0 & 0 & 0 & \frac{1}{\sqrt{3}} & 0 & \frac{1}{\sqrt{3}} & \frac{1}{\sqrt{3}}  & 0 \\
                   0 & 0 & 0 & 0 & 0 & 0 & 0 & 1\\
                   0 & \sqrt{\frac{2}{3}} & -\frac{1}{\sqrt{6}} & 0 & -\frac{1}{\sqrt{6}} & 0 & 0 & 0\\
                   0 & 0 & 0 & \frac{1}{\sqrt{6}} & 0 & \frac{1}{\sqrt{6}} & -\sqrt{\frac{2}{3}} & 0 \\
                   0 & 0 & -\frac{1}{\sqrt{2}} & 0 & \frac{1}{\sqrt{2}} & 0 &0 & 0\\
                   0 & 0 & 0 & -\frac{1}{\sqrt{2}} & 0 & \frac{1}{\sqrt{2}} & 0 & 0
\end{pmatrix} \label{eq:S3}
\end{align}

Next, we need a parameterized $P_3(\vec{\theta})$ rotation applied in the spin basis. In the parameterized gate we define a three-qubit unitary that acts on the two spin-$\frac{1}{2}$ spaces that come from the block diagonal decomposition of three qubits $\mathfrak{J}_\frac{1}{2}\otimes \mathfrak{J}_\frac{1}{2}\otimes \mathfrak{J}_\frac{1}{2}\simeq \mathfrak{J}_\frac{3}{2}\oplus \mathfrak{J}_\frac{1}{2}\oplus \mathfrak{J}_\frac{1}{2}$.
The difference between this gate and the one above is that the above two-qubit vertex gate lacks multiplicities, i.e., multiple blocks of the same size, meaning the only option is to have a phase on each different block. If we have multiple blocks of the same size, this indicates that there are multiple subspaces of the state space with the same total angular momentum and that multiple states exist with the same quantum numbers $\ket{J;J_z}$.
In terms of $\SU(2)$ equivariant maps, these are states that we can interchange without altering the structure of the space -- this implies that our vertex gates are not just phases on differing blocks but also unitaries that mix the multiple copies of $\ket{J;J_z}$ (see Fig.~\ref{fig:trees} for how our unitaries act on this space and Sec.~\ref{sec:rep_equivariant} for theoretical backgrounds). As an example, for our three-qubit space, we have one spin-$\frac{3}{2}$ space and two spin-$\frac{1}{2}$ spaces so it suffices to have a single unitary acting to mix the two $\ket{\frac{1}{2},J_z}$ states. The general matrix has the following form: 
\begin{align}\label{eq:3v}
\renewcommand\arraystretch{1.3}
    P_3(\vec{\theta}) = \begin{pmatrix}
        \begin{matrix}
          \Id_4  
        \end{matrix} &\rvline &
        \begin{matrix}
          0_4
        \end{matrix} \\
        \hline
        \begin{matrix}
          0_4
        \end{matrix} &\rvline &
        \begin{matrix}
          U_2(\vec{\theta}) \otimes \Id_2
        \end{matrix}
    \end{pmatrix} = \begin{pmatrix}
        \Id_2 &\rvline &0_2 \\ \hline
        0_2  &\rvline &U_2(\vec{\theta})
    \end{pmatrix} \otimes \Id_2
\end{align}
where $U_2(\vec{\theta})$ is a unitary matrix of dimension two, implying this gate has four real parameters. One might imagine that there could be a relative phase here on the isolated spin-$\frac{3}{2}$ space but (up to a global phase) this is a sub-case of the unitary acting on the two spin-$\frac{1}{2}$ components.
We note that this gate can be written as the \textsf{ControlledUnitary} gate between the first and second qubits (and acting trivially on the third qubit), which is generated by $\{ \ket{1}\bra{1}\otimes \Id_2, \ket{1}\bra{1}\otimes X, \ket{1}\bra{1}\otimes Y, \ket{1}\bra{1}\otimes Z \}$.

This leads to the three-qubit vertex gate definition.

\begin{definition}
The three-qubit vertex gate is composed as follows:
\begin{equation}
\begin{quantikz}
& \gate[3]{V_3(\vec{\theta})} & \qw \\
& & \qw \\
& & \qw
\end{quantikz} =
\begin{quantikz}
& \gate[3]{S_3} & \gate[3]{P_3(\vec{\theta})} & \gate[3]{S_3^\dagger} & \qw \\
&               & & & \qw\\
&               & & & \qw
\end{quantikz}
\end{equation}
where $S_3$ is the three qubit Schur gate and $P_{3}(\vec{\theta})$ is the controlled unitary seen in Eq.~\eqref{eq:3v}.
\end{definition}

Our construction extends to arbitrary $k$-qubit gates. 
In general, these spin-network circuits have the following shape:

\begin{equation}
V_k(\vec{\theta}) = S_k P_k(\vec{\theta}) S_{k}^\dagger
\end{equation}

Here, $\vec{\theta}$ is the vector of trainable parameters.
These are the free variables needed to parameterize the space of the $l$ different irreps that make up the spin basis of $k$ qubits $\oplus_{i=1}^l (U_{i} \otimes \Id_{ d_i})$ where each $U_i \in \mathrm{U} (m_i)$ is unitary of the size of the multiplicity of the $i^{th}$ representation and $d_i$ is the dimension of the $i^{th}$ irrep (i.e., $2J+1$ where $J$ is the spin number of the subspace).
These unitaries mix the states with the same $J_z$ value between the repeated irreps (again see Sec.~\ref{sec:rep_equivariant}).

Moreover, it is also not difficult to find a decomposition of $V_k(\vec{\theta})$ into an elementary gate set, e.g., single-qubit rotation and CNOT gates.
Specifically, the decomposition of the Schur gate, $S_k$, is extensively studied in Refs.~\cite{bacon2006efficient,Krovi2019efficienthigh}.
Implementing an $U(2^k)$ gate acting on the subspace of $n$-qubit system is also well known, which could be found in the standard textbook~\cite{NC10th} (see also the original reference~\cite{barenco1995elementary}).
Thus, for any $k$, one can decompose our vertex gates into single and two-qubit gates and further into Clifford+$T$ gates when implemented in a fault-tolerant quantum computer.

For example, when employing the standard universal gate set consisting of single-qubit Pauli rotation gates ($\mathsf{RX}$, $\mathsf{RY}$, and $\mathsf{RZ}$) and the $\mathsf{CNOT}$ gate,
the two-qubit Schur gate, $S_2$, can be decomposed into $15$ single-qubit Pauli rotation, and $3$ $\mathsf{CNOT}$ gates.
Since $P_2(\theta)$ corresponds to a controlled-phase shift gate, we can implement it using three $\mathsf{RZ}$ and two $\mathsf{CNOT}$ gates.
In total, the two-qubit vertex gate can be decomposed into $8$ $\mathsf{CNOT}$ gates and $33$ single-qubit Pauli rotations.
Likewise, one can implement the three-qubit vertex gate using $48$ $\mathsf{CNOT}$ and $159$ single-qubit rotation gates.
These numbers correspond to the most elementary, error-free decompositions; more efficient implementations may be achieved when small error tolerances and ancilla qubits are permitted.

An interesting question is how the few-qubit gates introduced in this section act on the global $\SU(2)$ subspace.
For example, let us consider a spin-3 irreducible subspace of 8 qubits (e.g., a state $\cos(\theta) |11111110 \rangle + \sin(\theta) |11011111 \rangle$ lives in this subspace).
How can we write down the matrix form of the gate in this subspace?
In the following section, we answer this question by outlining the theory of $\SU(2)$ equivariant gates from a global perspective. 
Interestingly, we will show that all $\SU(2)$ equivariant gates are the generalized permutations introduced in Ref.~\cite{zheng2022super}.

\section{Equivariant gates from representation theory}\label{sec:rep_equivariant}
In the previous section, we have introduced the Schur map for constructing gates that commute with the $\SU(2)$ group action.
However, the transformed basis from the Schur map only block diagonalizes $\SU(2)$ action, and an additional parameterized unitary gate (introduced as $P(\theta)$) acting between the blocks was necessary to build an equivariant gate.
In this section, we completely characterize all possible forms of such unitary gates by developing a general theory of $\SU(2)$ equivariant operations.
Furthermore, using the representation theory of $\SU(2)$ and the duality between the permutation group $\Sym_n$ and $\SU(2)$,
we prove that $\SU(2)$ equivariant operations are generalized permutations (which we formally define below), and conversely, all generalized permutations are also equivariant operators.
Using this result, we prove that our construction of equivariant gates gives the identical set of gates from the twirling formula and parameterized permutations introduced in Refs.~\cite{zheng2022super,meyer2023exploiting}.
We further answer the question raised at the end of the previous section using this identification.

\subsection{Equivariant operations as the commutant algebra of a representation}

Let us start with the definition of the commutant algebra.
\begin{definition}\label{def:com}
For a given representation $R:G \rightarrow \mathrm{GL}(\mathbb{C}^n)$, we define the commutant algebra $C(R)$ as
\begin{align}
    C(R) = \{T \in \mathcal{M}_n(\mathbb{C}): TR(g) = R(g)T \text{ for all } g \in G \},
\end{align}
where $\mathcal{M}_n(\mathbb{C})$ is the set of $n\times n$ complex matrices.
\end{definition}
One can verify that $C(R)$ forms an algebra (under matrix addition and multiplication).
This tells us that equivariant gates for $U^{\otimes n}$ with $U \in \mathrm{SU}(2)$ are nothing but unitary operators in $C(U^{\otimes n})$.

Throughout the rest of this subsection, we will construct a complete set of equivariant gates. To achieve this, it will be practical to pay closer attention to the structure of the commutant algebra. To this end, we consider the following lemmas.
\begin{lemma}[Schur's lemma]\label{lem:schur}
    A homomorphism preserving the group structure $f \in \mathrm{Hom}_G(V,W)$ is a homomorphism satisfying $f(gv) = g f(v)$ for all $g \in G$ and $v \in V$. 
    If $V$ and $W$ are two irreducible representations of a group $G$ over $\mathbb{C}$, then $f$ must be $c\Id$ for $c \in \mathbb{C}$ or $0$.
\end{lemma}

In short, a structure-preserving map between two irreps is either proportional to the identity (which implies that the vector space $V$ and $W$ are essentially the same) or zero (they are different irreps). A proof can be found in Refs.~\cite{sagan2013symmetric,hall2013lie}.
As $T$ in Definition~\ref{def:com} is a linear map, the condition $TR(g) = R(g)T$ implies that $T \in \mathrm{Hom}_G(\mathbb{C}^n,\mathbb{C}^n)$.
From this, we can more easily construct the commutant algebra for some simple cases.
For example, the commutant of a direct sum of differing irreps is a direct sum of two scaled identity maps. 

\begin{lemma}
Let $R^{(1)}$ and $R^{(2)}$ be different irreducible representations of a group $G$ with dimensions $d_1$ and $d_2$, respectively.
Let us consider a representation $R = R^{(1)} \oplus R^{(2)}$, written as
\begin{align}
    R(g) = \begin{pmatrix}
        R^{(1)}(g) & 0 \\
        0 & R^{(2)}(g)
    \end{pmatrix}.
\end{align}
Then we have
\begin{align}
    C(R) = \{c_1 \Id_{d_1} \oplus c_2 \Id_{d_2} : c_1, c_2 \in \mathbb{C} \}.
\end{align}
\end{lemma}
\begin{proof}
Let $T$ be a matrix with internal blocks $T_{1,1}, T_{1,2}, T_{2,1}, T_{2,2}$ given by
\begin{align}
    T = \begin{pmatrix}
        T_{1,1} & T_{1,2} \\
        T_{2,1} & T_{2,2}
    \end{pmatrix}.
\end{align}
If $TR = RT$, 
\begin{align*}
    &T_{1,1} R^{(1)} = R^{(1)} T_{1,1},\qquad T_{1,2} R^{(2)} = R^{(1)} T_{1,2}, \\
    &T_{2,1} R^{(1)} = R^{(2)} T_{2,1},\qquad T_{2,2} R^{(2)} = R^{(2)} T_{2,2}.
\end{align*}
Using Schur's lemma, we obtain $T_{1,1}=c_1 \Id$, $T_{2,2}=c_2 \Id$, $T_{1,2}=T_{2,1}=0$.
\end{proof}

The situation is more complicated in cases where we have a direct sum of the same representation. In this case, the commutant is not simply a direct sum but allows for mixing between the irreps. As we see further below, this will correspond to mixing between elements of the repeated irreps, which are the same.

\begin{lemma}
    We now consider a direct sum of the same representation $R=R^{(1)}\oplus R^{(1)}$. Then we have
    \begin{align}
        C(R) = \mathcal{M}_{2}(\mathbb{C}) \otimes \Id_{d_1}.
    \end{align}
\end{lemma}
\begin{proof}
    As before, we write $T \in C(R)$ in a block-diagonal matrix. Then $TR=RT$ gives 
    \begin{align}
        T_{i,j} R^{(1)} = R^{(1)} T_{i,j}.
    \end{align}
    Schur's lemma implies that each $T_{i,j}$ is proportional to $\Id$, i.e., $T_{i,j} = c_{i,j} \Id$ for $c_{i,j} \in \mathbb{C}$.
    Thus we have        
    \begin{align}
        T = \begin{pmatrix}
            c_{1,1} \Id & c_{1, 2} \Id \\
            c_{2,1} \Id & c_{2,2 }\Id
        \end{pmatrix} = \begin{pmatrix}
            c_{1,1} & c_{1,2} \\
            c_{2,1} & c_{2,2}
        \end{pmatrix} \otimes \Id.
    \end{align}
\end{proof}

Now let us generalize the above results. Let $R$ be a representation of $G$ on $V$. Then Maschke's theorem (for finite groups) or the Peter–Weyl Theorem (for Lie groups) asserts that $V$ is decomposable into a direct sum of irreducible representations
\begin{align}
    V \simeq m_1 R^{(1)} \oplus m_2 R^{(2)} \oplus \cdots \oplus m_k R^{(k)}, \label{eq:maschke_irrep_decomp}
\end{align}
where $m R = R \oplus R \oplus \cdots \oplus R$ signifies $m$ repetitions of the same representation, and $\{R^{(i)}\}$ are the different irreducible representations.
Applying the above lemmas gives the following theorem.
\begin{theorem}
Under the decomposition given by Eq.~\eqref{eq:maschke_irrep_decomp}, the commutant is given by
\begin{align}
    C(R) = \{\oplus_{i=1}^k (M_{i} \otimes \Id_{d_i}): M_{i} \in \mathcal{M}_{m_i} (\mathbb{C}) \text{ for all } i\} \label{eq:commutant}
\end{align}
where each $d_i$ is the dimension of the representation $R^{(i)}$.
\end{theorem}
Given that a square matrix $M \oplus N$ is unitary iff $M$ and $N$ are both unitary matrices, we obtain the following corollary.

\begin{corollary}\label{col:unitary_commutator}
All unitary operators commuting with $R$ are given by
\begin{align}
    C(R) \cap \mathrm{U}(d) = \{\oplus_{i=1}^k (U_{i} \otimes \Id_{ d_i}): U_{i} \in \mathrm{U} (m_i) \text{ for all } i\}, \label{eq:commutant_unitary}
\end{align}
where $d = \mathrm{dim}V = \sum_{i=1}^k m_i d_i$ is the dimension of $V$.
\end{corollary}

The Corollary tells us the exact form of intermediate unitary gates $P(\theta)$ we should use for $\SU(2)$ equivariant gates, which is evident from the following example.

\begin{example}
For a system with three qubits, we can decompose the space under $\SU(2)$ as
\begin{align}
    (\mathfrak{J}_{1/2})^{\otimes 3} \simeq \mathfrak{J}_{3/2} \oplus \mathfrak{J}_{1/2} \oplus \mathfrak{J}_{1/2},
\end{align}
where $\mathfrak{J}_s$ is a space of total spin $s$ with dimension $2s+1$. Note that the basis transformation from the computational basis to the total spin basis is nothing but the Schur transformation given in the previous section [Eq.~\ref{eq:S3}].
We can now see that the unitary operators that commute with $\SU(2)$ are given (up to a global phase) by 
\begin{align}
    \begin{pmatrix}
        \begin{matrix}
          \Id_4  
        \end{matrix} &\rvline &
        \begin{matrix}
          0_4
        \end{matrix}\\
        \hline
        \begin{matrix}
          0_4
        \end{matrix} &\rvline &
        \begin{matrix}
          U_2 \otimes \Id_2
        \end{matrix}\\
    \end{pmatrix},
\end{align}
which is the gate we defined in the previous section.
\end{example}

\subsection{$\mathrm{SU}(2)$ equivariant gates are generalized permutations}
We now completely characterize $\mathrm{SU}(2)$ equivariant gates for $n$ qubits using the above results by computing the multiplicity of each representation.
Our main tool is the Schur-Weyl duality, which posits the duality between the irreducible representation of the symmetric group $\Sym_n$ and that of $\SU(2)$. Thus, the multiplicity is given by the dimension of the corresponding irreducible representation of $\Sym_n$.

Let us first define two group actions.
For $U \in \mathrm{SU}(2)$, we define its action on $(\mathbb{C}^2)^{\otimes n}$ as 
\begin{align}
    U (\ket{v_1} \otimes \ket{v_2} \otimes \cdots \otimes \ket{v_n}) = \ket{U v_1} \otimes \ket{U v_2} \otimes \cdots \ket{U v_n},
\end{align}
where each $v_i$ is a vector in $\mathbb{C}^2$.
In matrix form, this action is nothing but $U^{\otimes n}$.

Another group we consider is the symmetric group $\Sym_n$. For $\alpha \in \Sym_n$, we define 
\begin{align}
    &\alpha (\ket{v_1} \otimes \ket{v_2} \otimes \cdots \otimes \ket{v_n}) \nonumber \\
    & \quad = \ket{v_{\alpha^{-1}(1)}} \otimes \ket{v_{\alpha^{-1}(2)}} \otimes \cdots \otimes \ket{v_{\alpha^{-1}(n)}}.
\end{align}
We can also write down a matrix representation of this group action.
Let us consider a transposition $\tau =(a,b) \in \Sym_n$ first, which just swaps the $a$-th and $b$-th qubit. In matrix form, this operation is written as
\begin{align}
    \tau = \frac{1}{2} \pmb{\sigma}^a \cdot \pmb{\sigma}^b + \frac{1}{2} \Id, \label{eq:spin_sum_transposition}
\end{align}
where $\pmb{\sigma}^i=\{\sigma^i_{x}, \sigma^i_{y}, \sigma^i_{z}\}$ is a vector of Pauli matrices acting on the $i$-th qubit. As any permutation $\alpha$ in $\Sym_n$ can be decomposed into transpositions, i.e., $\alpha = \tau_k \cdots \tau_2 \tau_1$ where each $\tau_i = (a_i, b_i)$ is a transposition, we obtain
\begin{align}
    \alpha &= \bigl(\frac{1}{2} \pmb{\sigma}^{a_k} \cdot \pmb{\sigma}^{b_k} + \frac{1}{2} \Id \bigr) \cdots \bigl(\frac{1}{2} \pmb{\sigma}^{a_2} \cdot \pmb{\sigma}^{b_2} + \frac{1}{2} \Id \bigr) \nonumber \\
    &\qquad \bigl(\frac{1}{2} \pmb{\sigma}^{a_1} \cdot \pmb{\sigma}^{b_1} + \frac{1}{2} \Id \bigr).
\end{align}

A crucial property of these two group actions is that they commute with each other, i.e., $U \alpha = \alpha U$.
One can easily check this for a product state
\begin{align*}
    U \alpha (\ket{v_1} \otimes \cdots \otimes \ket{v_n}) &=  U (\ket{v_{\alpha^{-1}(1)}} \otimes \cdots \otimes \ket{v_{\alpha^{-1}(n)}}) \\
    &=\ket{U v_{\alpha^{-1}(1)}} \otimes \cdots \otimes \ket{U v_{\alpha^{-1}(n)}} \\
    &=\alpha (\ket{U v_1} \otimes \cdots \otimes \ket{U v_n}) \\ 
    &=\alpha U (\ket{v_1} \otimes \cdots \otimes \ket{v_n}),
\end{align*}
which can be extended linearly to all vectors in the space.
Thus, it follows that a permutation is an $\mathrm{SU}(2)$ equivariant operation.
This fact is also the basis of the Schur-Weyl duality we introduce below.

Inspired by Ref.~\cite{zheng2023speeding}, we further consider an operator
\begin{align}
    Q = e^{\sum_{i=1}^k c_i \alpha_i} = \sum_{l=0}^\infty \frac{1}{l!} (\sum_{i=1}^k c_i \alpha_i)^l \in \mathbb{C}[\Sym_n],
\end{align}
where $c_i \in \mathbb{C}$ and $\alpha_i \in \Sym_n$, which we call generalized permutations.
Here, $\mathbb{C}[\Sym_n]$ is a set of all linear combinations of $\sigma \in \Sym_n$ with complex-valued coefficients.
From the expansion, we see that $Q$ also commutes with $U \in \mathrm{SU}(2)$, which implies that $Q$ is an $\mathrm{SU}(2)$ equivariant operation as well (albeit not unitary, in general).
If we further restrict unitarity, i.e., an operator $e^{\sum_i c_i \alpha_i}$ with anti-Hermitian $\sum_i c_i \alpha_i$, such an operator is an element of the set given by Eq.~\eqref{eq:commutant_unitary}.

We now prove the converse of the above statement, which is the main result of this section: All $\mathrm{SU}(2)$ equivariant unitary operators can also be written as a form of $\exp[\sum_{i=1}^k c_i \alpha_i]$.
Even though this can be understood as a consequence of von Neumann's double commutation theorem (see, e.g., Ref.~\cite{roberts2017chaos}), here we provide constructive proof with a concrete example.
The first ingredient for the proof is the \textit{Schur-Weyl duality}.
\begin{theorem}[Schur-Weyl duality]\label{thm:schur-weyl}
    Under the group actions of $U \in \mathrm{SU}(2)$ and the symmetric group $\alpha \in \Sym_n$, the tensor-product space decomposes into a direct sum of tensor products of irreducible spaces that determine each other. Precisely, we can write
    \begin{align}
        (\mathbb{C}^2)^{\otimes n} \simeq \bigoplus_D  \pi^D_n \otimes \mathfrak{J}_{D}
    \end{align}
    where the summation is over the Young diagram $D$ with $n$ boxes and at most two rows.
    For each $D$ with $r_1$ boxes in the first row and $r_2$ boxes in the second row, $\mathfrak{J}_{D}$ is the irreducible representation of $\SU(2)$ with total spin $J=(r_1-r_2)/2$, and $\pi^D_n$ is the irreducible representation of the symmetric group associated with the given Young diagram $D$.
\end{theorem}

We formally introduce the Young diagram and the irreducible representation of $\Sym_n$ in Appendix~\ref{app:rep_sym}.
However, for the rest of the discussion in this section, it is fine to skip the details and only consider the dimension of $\pi_{n}^D$, as we show in the following Corollary.

\begin{corollary} \label{col:n_qubit_irrep_decomp}
From the Schur-Weyl duality, one obtains
\begin{align}
    (\mathbb{C}^2)^{\otimes n} \simeq \bigoplus_{i=0}^{\lfloor n/2 \rfloor } m_i \mathfrak{J}_{s_i}
\end{align}
where $m_i$ is the dimension of the irreducible representation of $\Sym_n$  whose Young diagram $D_i$ has $n-i$ boxes in the first row and $i$ boxes in the second row, and $s_i = n/2-i$ is the total spin.

The dimension of the irreducible representation can be computed using the Hook length formula. After some steps (see, e.g., Corollary 3.8 in Ref.~\cite{watts2024relaxations}), one can obtain
\begin{align}
    m_i = \begin{cases}
    1, & \text{if } i=0, \\
    \binom{n}{i} - \binom{n}{i-1}, & \text{otherwise}.
    \end{cases}
\end{align}
\end{corollary}

We then apply Corollary~\ref{col:unitary_commutator} to this decomposition and obtain all possible $\SU(2)$ equivariant gates, given by
\begin{align}
    U = \Bigl\{ \bigoplus_{i=0}^{\lfloor n/2 \rfloor} (U_{i} \otimes \Id_{d_i}): U_i \in U(m_i) \Bigr\}.
\end{align}
In addition, as each $U(m_i)$ has $m_i^2$ independent generators, the total number of parameters is given by
\begin{align}
    \sum_{i=0}^{\lfloor n/2 \rfloor } m_i^2 = 
    \frac{1}{n+1} {2n \choose n}.
\end{align}
Note that Ref.~\cite{nguyen2024theory} also presents the same result.
We also note that, for a quantum gate, we can subtract one from this formula as there is a redundancy for the global phase.

Another ingredient we need is the completeness of the irreducible representation.
\begin{theorem}[The density theorem~\cite{jacobson1945structure}]
    Let $V=\mathbb{C}^n$ be an irreducible finite-dimensional representation of a group $G$, i.e., there is a map $R: G \rightarrow \mathrm{GL}(\mathbb{C}^d)$.
    Then $\{R(g): g \in G\}$ spans $\mathcal{M}_d(\mathbb{C})$.
\end{theorem}

See, e.g., Ref.~\cite{etingof2011introduction} for a proof. The theorem implies that for any $M \in \mathcal{M}_d(\mathbb{C})$, we can find $g_i \in G$ and $c_i \in \mathbb{C}$ such that $M = \sum_{i=1}^k c_i R(g_i)$ when $\mathbb{C}^{d}$ is the irreducible representation of $G$.

Using the Schur-Weyl duality and the density theorem, we now prove the equivalence between a generalized permutation group action and $\mathrm{SU}(2)$ equivariant unitary gates.

\begin{theorem}\label{thm:eq-u}
    For any $\mathrm{SU}(2)$ equivariant unitary gate $T$, we can find $c_i \in \mathbb{C}$ and $\alpha_i \in \Sym_n$ such that
    \begin{align}
        T = e^{\sum_{i=1}^k c_i \alpha_i}. \label{eq:exp_r_sym}
    \end{align}
\end{theorem}
\begin{proof}
    First, from Corollary~\ref{col:n_qubit_irrep_decomp}, we obtain
    \begin{align}
        (\mathbb{C}^2)^{\otimes n} \simeq \bigoplus_{i=0}^{\lfloor N/2 \rfloor } m_i \mathfrak{J}_{s_i}.
    \end{align}
    
    Then let $H$ be the generator of $T$, i.e., $T = e^{iH}$ and $H$ is a Hermitian matrix. Looking at Corollary~\ref{col:unitary_commutator}, we can move from the description of equivariant unitaries to their generators and see that $H$ can be written as 
    \begin{align}
        H = \bigoplus_i h_i \otimes \Id_{2s_i + 1} = \sum_i h_i P_i \label{eq:hermitian_decomposition}
    \end{align}
    where $h_i$ is a hermitian matrix in $\mathcal{M}_{m_i}(\mathbb{C})$ and $P_i$ is a projector onto a subspace with total spin $2s_i + 1$.
    From the density theorem, one can find $\{c_{ij} \in \mathbb{R}\}$ and $\{\alpha_{ij} \in \Sym_n \}$ such that $h_i = \sum_j c_{ij} \alpha_{ij}$ for each $i$.
    Moreover, each projector $P_i$ can be written as
    \begin{align}
        P_i = \prod_{j \neq i} \frac{J^2 - s_j(s_j+1)}{s_i(s_i+1) - s_j (s_j+1)},
    \end{align}
    where $\pmb{J} = \sum_{i=1}^n \pmb{\sigma}^i/2$ is the total spin operator and $J^2 = \pmb{J} \cdot \pmb{J}$.
    As $J^2$ has eigenvalues $s_i(s_i+1)$ for each subspace $\mathfrak{J}_{s_i}$, one can verify that the given operator is indeed a projector.
    After rewriting
    \begin{align}
        J^2 = \frac{1}{4} \bigl( 3n + \sum_{i \neq j} \pmb{\sigma}^i \cdot \pmb{\sigma}^j \bigr) = \frac{4n-n^2}{4} + \sum_{i > j} (i,j)
    \end{align}
    where $(i,j)$ is a transposition, we see that $J^2 \in \mathbb{R}[\Sym_n]$ where $\mathbb{R}[\Sym_n]$ is the set of all linear combination of elements in $\Sym_n$ with real coefficients. If we again look at Eq.~\eqref{eq:hermitian_decomposition}, 
    we can now see that as $h_i\in \mathbb{C}[\Sym_n]$ our unitary $T=e^{iH}$ is indeed an exponentiated sum of permutations with coefficients in $\mathbb{C}$~\footnote{Note that $h_i$ is not necessarily in $\mathbb{R}[\Sym_n]$ since a permutation $\alpha \in \Sym_n$ can be non-Hermitian.}.
\end{proof}

This theorem provides the theoretical foundation for several prior works~\cite{nguyen2024theory,zheng2022super,zheng2023speeding}. Nevertheless, as discussed earlier, implementing a gate of the form in Eq.~\eqref{eq:exp_r_sym} directly on a quantum circuit is nontrivial. By contrast, the vertex gates introduced in the preceding section admit a straightforward implementation, and Corollary~\ref{col:unitary_commutator} shows that any $\SU(2)$-equivariant gate can be realized as a vertex gate.

\subsection{Twirling and permutations}
In Ref.~\cite{meyer2023exploiting}, the twirling method is proposed to construct an equivariant unitary gate.
For a given Hermitian matrix $H$ that is the generator of a unitary gate $V=\exp(iH)$ and a Lie group $\mathcal{G}$, one obtains an equivariant version of it using the twirling formula:
\begin{align}
    \mathcal{T}_U[H] = \int d \mu(g) R(g) H R(g)^\dagger,
\end{align}
where $\mu(g)$ is the Haar measure for the Lie group $\mathcal{G}$. 
Then $\mathcal{T}_U[H]$ commutes with any $h \in \mathcal{G}$ due to a defining property of the Haar measure, and so does the gate $\exp\{ i\mathcal{T}_U[H] \}$.

We now show that the twirling formula yields a generalized permutation for $\mathcal{G} = \SU(2)$.
For a Hermitian matrix $H \in \mathcal{M}_{2^{n}}(\mathbb{C})$, we obtain
\begin{align*}
    \mathcal{T}_U[H] &= \int d \mu(g) R(g) H R(g)^\dagger \\
    &= \int_U dU U^{\otimes n} H (U^\dagger)^{\otimes n} \\
    &= \sum_{\sigma, \tau \in \Sym_n} \Wg(\sigma^{-1} \tau, d) \Tr[H \tau] \sigma, \numberthis
\end{align*}
where $d=2^n$ is the dimension of the Hilbert space, $\Wg(\sigma, d)$ is the Weingarten function, and we identified $\sigma,\tau \in \Sym_n$ as an operator using the representation (see e.g., Refs.~\cite{roberts2017chaos,brandao2021models} for the explanation how the last line is obtained). Ultimately, this is a permutation scaled by a real coefficient as required.
Furthermore, as $\mathcal{T}_U[H]$ is also Hermitian by definition, we know that $\mathcal{T}_U[H]$ is a Hermitian element of $\mathbb{C}[\Sym_n]$, which can be a generator for an equivariant unitary gate.

On the other hand, all generators of equivariant gates can be obtained from the twirling formula.
In the spin-basis, we know that each generator of an equivariant gate is given by Eq.~\eqref{eq:hermitian_decomposition}, i.e., $H \simeq \oplus_i h_i \otimes \Id_{d_i}$ (where the dimension of $h_i$ and $d_i$ are obtained from the Schur-Weyl duality). 
As this is an element of the commutant [Eq.~\eqref{eq:commutant}], $H$ is also equivariant, i.e., $H U^{\otimes N} = U^{\otimes N} H$, so $\mathcal{T}_U[H] = H$.
In other words, the set of all generators of equivariant gates and the set of all twirled generators are the same:
\begin{align*}
    &\bigl\{H \in \mathcal{M}_{2^{n}}(\mathbb{C}) : U^{\otimes n} e^{iH} = e^{iH}U^{\otimes n} \\
    & \qquad \text{ for all } U \in \SU(2) \text{ and } H=H^\dagger \bigr\} \\
    &= \bigl\{ \mathcal{T}_U[H] : H \in \mathcal{M}_{2^{n}}(\mathbb{C}) \text{ and } H^\dagger = H\bigr\} \numberthis.
\end{align*}

\subsection{Revisiting three-qubit $\mathrm{SU}(2)$ equivariant gates}
In this subsection, using the three-qubit vertex gate as an example, we illustrate how to represent our equivariant gates as elements of $\mathbb{C}[\Sym_n]$.
We apply Theorem~\ref{thm:eq-u} to the three-qubit gate we have found in Sec.~\ref{sec:spin-network-circuits}, using the Schur map given in Eq.~\eqref{eq:S3}.
A direct consequence of the Schur transform is that it defines invariant subspaces under $U^{\otimes 3}$ for any $U \in \SU(2)$, given by $\mathfrak{J}_{3/2} = \mathrm{span}\{S_3^\dagger \ket{0}, S_3^\dagger \ket{1}, S_3^\dagger \ket{2}, S_3^\dagger \ket{3}\}$, $\mathfrak{J}_{1/2}^a=\mathrm{span}\{S_3^\dagger \ket{4}, S_3^\dagger \ket{5}\}$, and $\mathfrak{J}_{1/2}^b=\mathrm{span}\{S_3^\dagger\ket{6}, S_3^\dagger \ket{7}\}$ where the superscripts $a,b$ are used to distinguish two spin-1/2 spaces.
From the structure of $P(\vec{\theta})$, we know the gate has four generators given by $\{G_I := \mathbf{0}_4 \oplus \Id_4, G_X := \mathbf{0}_4 \oplus(X\otimes \Id_2), G_Y := \mathbf{0}_4 \oplus (Y\otimes \Id_2), G_Z := \mathbf{0}_4 \oplus (Z\otimes \Id_2) \}$, where $\mathbf{0}_4$ acts on $\mathfrak{J}_{3/2}$ whereas $X,Y,Z$ mixes $\mathfrak{J}_{1/2}^a$ and $\mathfrak{J}_{1/2}^b$.
One can also see that a permutation in $\Sym_3$ mixes subspaces $\mathfrak{J}_{1/2}^a$ and $\mathfrak{J}_{1/2}^b$ (whereas it acts trivially on $\mathfrak{J}_{3/2}$ subspace).

A matrix representation of a permutation for $\{\mathfrak{J}_{1/2}^a, \mathfrak{J}_{1/2}^b\}$ is obtained by applying each permutation to a basis vector, which is given as
\begin{align}
    (1,2) &= \begin{pmatrix}
        1 & 0 \\
        0 & -1
    \end{pmatrix} \otimes \Id_2 = Z \otimes \Id_2\\ 
    (2,3) &= \begin{pmatrix}
        -1/2 &  -\sqrt{3}/2 \\ 
        -\sqrt{3}/2 & 1/2
    \end{pmatrix} \otimes \Id_2 \nonumber \\
    & = -\frac{1}{2} Z \otimes \Id_2 - \frac{\sqrt{3}}{2} X \otimes \Id_2 \\
    (1,3) &= \begin{pmatrix}
        -1/2 &  \sqrt{3}/2 \\
        \sqrt{3}/2 &  1/2
    \end{pmatrix} \otimes \Id_2 \nonumber \\
    & = -\frac{1}{2} Z \otimes \Id_2 + \frac{\sqrt{3}}{2} X \otimes \Id_2,
\end{align}
Each matrix should be read as follows.
For example, if we apply $(2,3)$ to $S_3^\dagger \ket{4}$, we have
\begin{align}
    (2, 3) S_3^\dagger \ket{4} = -\frac{1}{2} S_3^\dagger \ket{4} - \frac{\sqrt{3}}{2} S_3^\dagger \ket{6},
\end{align}
where the coefficients are from the first column of the matrix representation of $(2,3)$.
Note that the permutation transforms $S_3^\dagger \ket{5}$ exactly the same way (but mixes $S_3^\dagger \ket{5}$ and $S_3^\dagger \ket{7}$).
Using the above expressions, the remaining elements are obtained as follows (where we dropped $\otimes \Id_2$ to simplify the notation):
\begin{align}
    (1,2,3) &= (1,2)(2,3) = - \frac{1}{2} \Id - i \frac{\sqrt{3}}{2}Y \\
    (1,3,2) &= (1,2)(1,3) = - \frac{1}{2} \Id + i \frac{\sqrt{3}}{2}Y. 
\end{align}

Thus we have 
\begin{align}
    I &= 1,\\
    X &= -\frac{2}{\sqrt{3}}[ (2,3) + 1/2 (1,2)] \\
    Y &= i\frac{1}{\sqrt{3}} [2 (1,2,3) + 1 ], \\
    Z &= (1,2).
\end{align}

However, these operators cannot be generators of our gate as they do not annihilate the $J=3/2$ subspace (recall that our generators have $\mathbf{0}_4$ on the $\mathfrak{J}_{3/2}$ subspace).
Thus, we need a projector to the $J=1/2$ subspace, which is given by
\begin{align}
    P_{J=1/2} = \frac{J^2 - 15/4}{3/4 - 15/4} = \frac{5}{4} - \frac{1}{3}J^2
\end{align}
where $J^2$ is
\begin{align}
    J^2 = \frac{1}{4}[\pmb{\sigma}^1 + \pmb{\sigma}^2  + \pmb{\sigma}^3]^2 = \frac{3}{4} + [(1,2) + (2,3) + (1,3)].
\end{align}

By combining the projector and expressions of Pauli operators in $J=1/2$ subspaces, we can write three generators as
\begin{align}
    &G_I = 1 - \frac{1}{3}[(1,2) + (2,3) + (1,3)] \label{eq:G_I} \\
    &G_X = \nonumber \\
    & -\frac{2}{\sqrt{3}}\Bigl[ -\frac{1}{2} + (2,3) + \frac{1}{2}(1,2)- \frac{1}{2}(1,2,3) - \frac{1}{2}(1,3,2) \Bigr] \\
    &G_Y = i \frac{1}{\sqrt{3}}\Bigl[ 1 + 2(1,2,3) - (1,2) - (2,3) - (1,3) \Bigr] \\
    & G_Z = (1,2) -\frac{1}{3} [ 1 + (1,3,2) + (1,2,3) ] \label{eq:G_Z}
\end{align}
One can check that each generator annihilates the $\mathfrak{J}_{3/2}$ subspace (e.g., $G_X \ket{000} = 0$), and acts like a Pauli gate between the $\mathfrak{J}_{1/2}^a$ and $\mathfrak{J}_{1/2}^b$ subspaces (e.g., $G_X S_3^\dagger \ket{5} = S_3^\dagger \ket{7} $).
Also note that, as there is a freedom in choosing two $J=1/2$ subspaces (any unitary mixtures between $\mathfrak{J}_{1/2}^a$, $\mathfrak{J}_{1/2}^b$ are also valid subspaces),
the exact form of generators $\{G_I, G_X, G_Y, G_Z\}$ depends on the specific choice of the Schur gate $S_3$ (which is from Eq.~\eqref{eq:S3} for our case).

To summarize, any $\SU(2)$ equivariant gate on the three qubit can be written as
\begin{align}
    V(\vec{\theta}) &= S_3^\dagger P(\vec{\theta}) S_3  \nonumber \\
    &= \exp\Bigl[i \bigl\{\theta_0 G_I +  \theta_1  G_X + \theta_2 G_Y + \theta_3 G_Z \bigr\} \Bigr],
\end{align}
which is a generalized permutatation from Eq.~(\ref{eq:G_I}-\ref{eq:G_Z}).

We now answer the question raised at the end of the previous section. 
If we apply our three-qubit gate to the 3rd, 4th, and 7th qubits among eight qubits, we first obtain its representation as a generalized permutation between them and apply it to basis vectors of global spin subspaces. For example, $G_X$ for those qubits is given as
\begin{align}
    &G_X^{(3,4,7)} = \nonumber \\
    & \quad -\frac{2}{\sqrt{3}}\Bigl[ -\frac{1}{2} + (4,7) + \frac{1}{2}(3,4) - \frac{1}{2}(3,4,7) - \frac{1}{2}(3,7,4) \Bigr].
\end{align}
Then, one can construct its matrix form in a certain subspace (e.g., one of the $\mathfrak{J}_2$ subspaces) by applying it to the basis vectors of the subspace. Then the gate $\exp[-i\theta G_X^{(3,4,7)}]$ can be reconstructed by applying the exponential map.

We finalize this section by introducing an alternative description of these generators using the scalar products. For three operator vectors $\pmb{\sigma}^1$, $\pmb{\sigma}^2$, $\pmb{\sigma}^3$, the only possible scalar operators (that are invariant under the group transformation) obtained from those operators are $\pmb{\sigma}^1 \cdot \pmb{\sigma}^2$, $\pmb{\sigma}^2 \cdot \pmb{\sigma}^3$, $\pmb{\sigma}^1 \cdot \pmb{\sigma}^3$, and $\pmb{\sigma}^1 \cdot (\pmb{\sigma}^2 \times \pmb{\sigma}^3)$ up to constant factors, where $A \times B$ is the cross product between two vectors.
Thus, another possible representation of a parameterized three-qubit equivariant gate is 
\begin{align}
    W &= \exp \bigl[i (\theta_{12} \pmb{\sigma}^1 \cdot \pmb{\sigma}^2 + \theta_{23} \pmb{\sigma}^2 \cdot \pmb{\sigma}^3 + \theta_{13} \pmb{\sigma}^1 \cdot \pmb{\sigma}^3) \nonumber \\
    &\qquad \qquad + i\phi \pmb{\sigma}^1 \cdot (\pmb{\sigma}^2 \times \pmb{\sigma}^3) \bigr].
\end{align}
Then, it can be shown that this gate is the same as $V(\vec{\theta})$ up to a global phase.

Using
\begin{align}
    (\pmb{\sigma}^1 \cdot \pmb{\sigma}^2) (\pmb{\sigma}^2 \cdot \pmb{\sigma}^3) = \pmb{\sigma}^1 \cdot \pmb{\sigma}^3 + i \pmb{\sigma}^1 \cdot ( \pmb{\sigma}^2 \times \pmb{\sigma}^3),
\end{align}
and Eq.~\eqref{eq:spin_sum_transposition}, we obtain
\begin{align*}
    2i \pmb{\sigma}^1 \cdot ( \pmb{\sigma}^2 \times \pmb{\sigma}^3) &= [\pmb{\sigma}^1 \cdot \pmb{\sigma}^2, \pmb{\sigma}^2 \cdot \pmb{\sigma}^3] \\
    &= [2(1,2) - 1, 2(2,3) - 1] \\
    &= 4(1,2,3) - 4(1,3,2). \numberthis
\end{align*}
In addition, we need another identity $P_{J=3/2}^2=P_{J=3/2}$, which gives
\begin{align}
    (1,2,3) + (1, 3, 2) = (1,2) + (2,3) + (1,3) - 1.
\end{align}
Note that this equality only implies that the LHS and RHS act the same on our vector space. Of course, they are different elements in $\mathbb{C}[\Sym_n]$.

Combining all these together, we can write each generator of $W$ in terms of $\{G_I, G_X, G_Y, G_Z\}$ as
\begin{align}
    \pmb{\sigma}^1 \cdot \pmb{\sigma}^2 &= 2(1,2) - 1 = 1 - 2G_I + 2G_Z \\
    \pmb{\sigma}^2 \cdot \pmb{\sigma}^3 &= 2(2,3) - 1 = 1 - 2 G_I - \sqrt{3} G_X - G_Z \\
    \pmb{\sigma}^3 \cdot \pmb{\sigma}^1 &= 2(1,3) - 1 = 1 - 2 G_I + \sqrt{3} G_X - G_Z \\
    \pmb{\sigma}^1 \cdot (\pmb{\sigma}^2 \times \pmb{\sigma}^3) &= -\frac{i}{2} [4(1,2,3) - 4(1,3,2)] = -2\sqrt{3}G_Y,
\end{align}
which implies that $W$ is just another parameterization of $V(\vec{\theta})$ (up to a global phase).

\begin{figure}
    \centering
    \includegraphics[width=0.85\linewidth]{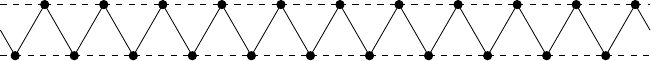}
    \caption{A one-dimensional triangular lattice. We solve the Heisenberg XXX model defined on this lattice using the equivariant gates. The interaction strength between qubits linked with solid lines is given by $\mathcal{J}_1$, whereas those between qubits linked with dashed lines are $\mathcal{J}_2$. }
    \label{fig:triangular-lattice}
\end{figure}

\section{Numerical Simulations}\label{sec:simulations}
In this section, we numerically demonstrate the efficacy of our equivariant gates for solving quantum many-body Hamiltonians. 
Our Hamiltonians are the Heisenberg XXX models (which are rotationally invariant) defined on frustrated lattices.
Even though the Heisenberg models are toy models, they play an important role in understanding the low-temperature physics of some exotic materials~\cite{shimizu2003spin}.
All numerical simulations in this section were performed using the PennyLane~\cite{bergholm2018pennylane} software package with the Lightning~\cite{Lightning} plugin.
Relevant source code is available in the GitHub repository~\cite{park2023_github_repo}.

\begin{figure}[b]
    \centering
    \includegraphics[width=0.98\linewidth]{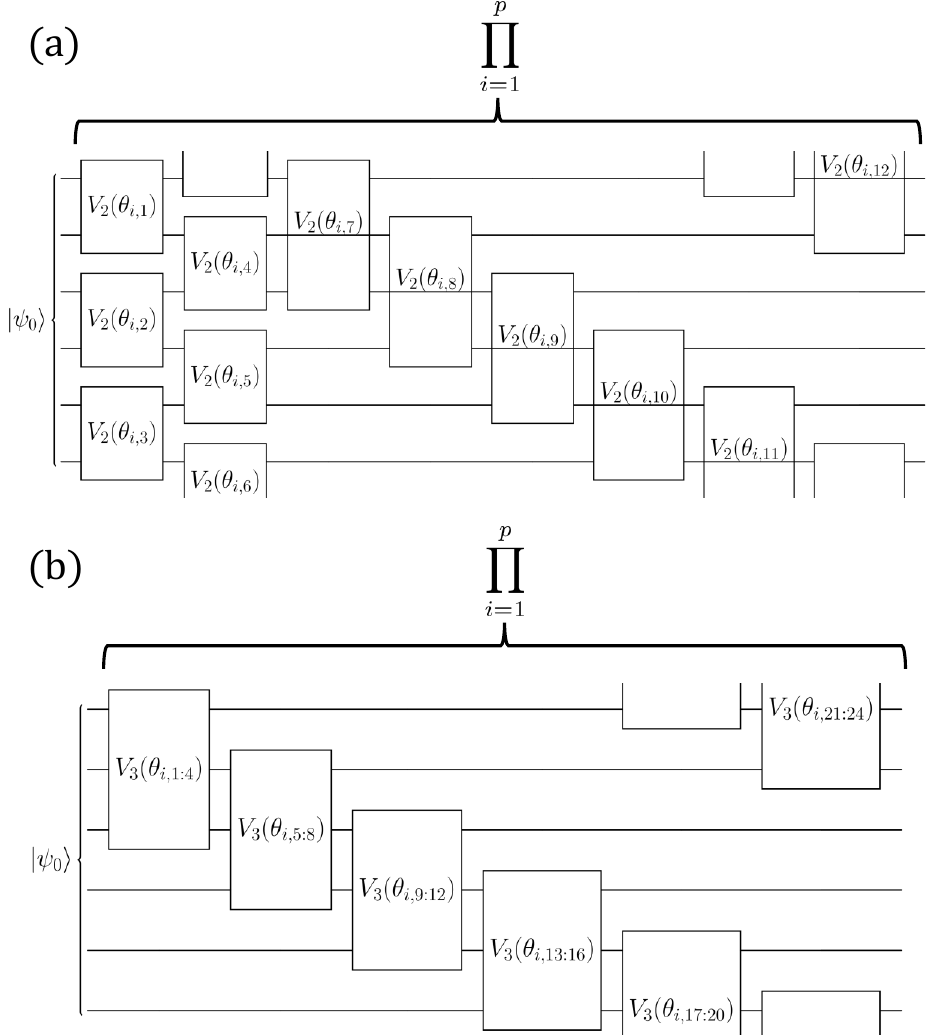}
    \caption{
    \label{fig:circuit_ansatze}
    Quantum circuit ans\"{a}tze for solving the Heisenberg XXX model on the triangular lattice with (a) two-qubit vertex gates and (b) three-qubit vertex gates. A wire passing through a gate indicates that the gate is not applied to that wire. 
    }
\end{figure}

\subsection{One-dimensional triangular lattice}
Let us first consider a one-dimensional triangular lattice as shown in Fig.~\ref{fig:triangular-lattice}.
The Hamiltonian we want to solve is 
\begin{align}
    H&= \mathcal{J}_1\sum_{i=1}^n \bigl[ \sigma_{i}^x \sigma_{i+1}^x + \sigma_{i}^y \sigma_{i+1}^y + \sigma_{i}^z \sigma_{i+1}^z \bigr] \nonumber \\
    & \quad + \mathcal{J}_2 \sum_{i=1}^n \bigl[ \sigma_{i}^x \sigma_{i+2}^x + \sigma_{i}^y \sigma_{i+2}^y + \sigma_{i}^z \sigma_{i+2}^z \bigr], \label{eq:ham_j1j2}
\end{align}
where we impose the periodic boundary condition $\sigma_{n+1}^{x,y,z} = \sigma_1^{x,y,z}$. Throughout the section, we fix $\mathcal{J}_1=1$ and consider $\mathcal{J}_2 \in \{0, 0.44\}$.

When $\mathcal{J}_2=0$, the Hamiltonian can be transformed into a stoquastic form~\cite{bravyi2006complexity} and a classical algorithm, the variational quantum Monte Carlo (vQMC) with a simple complex-valued restricted Boltzmann machine (RBM), can find the ground state energy extremely accurately~\cite{carleo2017solving}.
In contrast, such a transformation does not work for $\mathcal{J}_2 \gneq 0$~\cite{klassen2019two}, and the vQMC with the RBM deviates from the true ground state.
We here choose $\mathcal{J}_2 = 0.44$ as a recent study~\cite{park2022expressive} reported that such a deviation is maximized near this value.
Still, we note that the density matrix renormalization group can faithfully solve our model as the model is one-dimensional.

\begin{figure*}[t]
    \centering
    \includegraphics[width=0.85\linewidth]{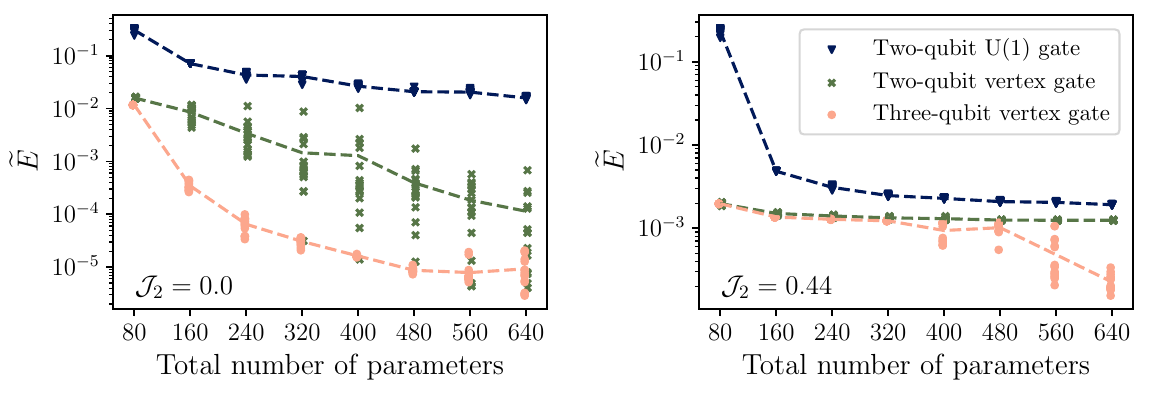}
    \caption{Normalized converged energies as functions of the total number of parameters in a given ansatz for $\mathcal{J}_2=0.0$ (left) and $\mathcal{J}_2 = 0.44$ (right). Each data point represents the converged energy obtained from an initial parameter.
    The dashed lines show the converged energies averaged over all instances for each circuit ansatz and the number of parameters.
    }
    \label{fig:one-dim-j1j2}
\end{figure*}

We compare the performance of three quantum circuit ans\"{a}tze for solving this Hamiltonian. 
The first ansatz utilizes the $U(1)$-equivariant gate that is introduced in Ref.~\cite{gard2020efficient}.
The gate is defined as
\begin{align}
    A(\theta, \phi) = \begin{pmatrix}
        1 & 0 & 0 & 0 \\
        0 & \cos \theta & e^{i \phi} \sin \theta & 0 \\
        0 & e^{-i \phi} \sin \theta &  -\cos \theta & 0 \\
        0 & 0 & 0 & 1
    \end{pmatrix}.
\end{align}
Using this gate, we defined a circuit
\begin{align}
    &| \psi(\{\theta_{ij}\}) \rangle = \prod_{i=p}^1 \Bigl[ \prod_{j=1}^{n} A_{j,j+2}(\theta_{i,2n+2j+1},\theta_{i,2n+2j}) \nonumber \\
    & \qquad \times \prod_{j=1}^{n/2} A_{2j,2j+1}(\theta_{i,n+2j+1},\theta_{i,n+2j}) \nonumber \\
    & \qquad \times \prod_{j=1}^{n/2} A_{2j-1,2j}(\theta_{i,2j+1},\theta_{i,2j}) \Bigr] \ket{\psi_0}, \label{eq:j1j2-u1-circuit}
\end{align}
where $A_{kl}$ is the gate $A$ defined above acting on $k$ and $l$-th qubits and $\ket{\psi_0} = \sqrt{2}^{\,-n/2}(\ket{01} - \ket{10})^{\otimes n/2}$ is a series of singlets.
While $\ket{\psi_0}$ is $\SU(2)$ invariant, since the circuit only preserves the $U(1)$ symmetry, the final state is not $\SU(2)$ invariant in general.
The total number of parameters of the circuit is $4np$.

Our second ansatz consists of the two-qubit vertex gates. The circuit is written as
\begin{align}
    | \psi(\{\theta_{ij}\}) \rangle &= \prod_{i=p}^1 \Bigl[ \prod_{j=1}^{n} V_{j,j+2}(\theta_{i,j+n}) \prod_{j=1}^{n/2} V_{2j,2j+1}(\theta_{i,j+n/2})\notag \\
    &\quad \times \prod_{j=1}^{n/2} V_{2j-1,2j}(\theta_{i,j}) \Bigr] \ket{\psi_0}, \label{eq:j1j2-two-qubit-vertex}
\end{align}
where $V_{kl}$ is the two-qubit vertex gate acting on $k$-th and $l$-th qubits.
As $\ket{\psi_0}$ is $\SU(2)$ invariant and our circuit is $\SU(2)$ equivariant, the output state is also $\SU(2)$ invariant.
The ansatz has a total of $2np$ parameters, where $p$ is the number of blocks in the ansatz.

Likewise, we also define the third ansatz that consists of the three-qubit vertex gates as
\begin{align}
    &| \psi(\{\theta_{i,j}\}) \rangle = \nonumber \\
    &\prod_{i=p}^1 \Bigl[ \prod_{j=1}^n V_{j,j+1,j+2} (\{\theta_{i,4j-3},\theta_{i,4j-2},\theta_{i,4j-1},\theta_{i,4j}\}) \Bigr] \ket{\psi_0},
\end{align}
where $V_{j,j+1,j+2}$ is the three-qubit vertex gate acting on qubits $\{j, j+1, j+2\}$. Also, recall that the three-qubit vertex gate has four parameters, so the ansatz has $4np$ parameters in total. 
Quantum circuit ans\"{a}tze for $n=8$ are explicitly shown in Fig.~\ref{fig:circuit_ansatze}.

We now solve the Hamiltonian from Eq.~\eqref{eq:ham_j1j2} with $n=20$ qubits for two different values of $\mathcal{J}_2 \in \{0.0, 0.44\}$ using the proposed ans\"{a}tze by simulating variational quantum eigensolvers (VQEs) using a classical simulator. 
For each ansatz, we optimize the parameters by minimizing $\langle H \rangle$ using the Adam optimizer.
For the hyperparameters $\beta_1=0.9$, $\beta_2=0.99$, $\epsilon=10^{-8}$, which are the default values defined in PennyLane~\cite{bergholm2018pennylane}, we run the Adam optimizer for different learning rates $\eta \in [10^{-3}, 2\times 10^{-3}, 5 \times 10^{-3}, 10^{-2}, 2\times 10^{-2}]$ using the circuit with two-qubit vertex gates and the depth $p=4$.
Among the tested learning rates, we choose $\eta= 5 \times 10^{-3}$ since it outperformed the others.

After optimization, we compute the converged normalized energies $\widetilde{E} = (\langle H \rangle - E_{\rm GS})/|E_{\rm GS}|$ where $E_{\rm GS}$ is the true ground state energy obtained from exact diagonalization.
We use the number of blocks $p=[2,4,6,8,10,12,14,16]$ for the ansatz with two-qubit vertex gates.
On the other hand, $p=[1,2,3,4,5,6,7,8]$ is used for the ansatz with three-qubit vertex gates.
In addition, inspired by Ref.~\cite{park2024hamiltonian}, we initialize the parameters using samples from the distribution $\mathcal{U}_{[0,\alpha]}$ divided by the total number of parameters where $\mathcal{U}_{[0,\alpha]}$ is the uniform distribution between $0$ and $\alpha$, and $\alpha$ is a hyperparameter giving a relative scaling.
For the simulation, we choose $\alpha=1$, which shows relatively better results than other values tested.
We also note that our simulation is performed by computing exact gradients (without shot noise), which is more efficient for classical simulators.

For $16$ random initial parameters, we plot the converged normalized energies in Fig.~\ref{fig:one-dim-j1j2} as a function of the total number of parameters.
We observe several interesting points.
First, the circuit with the two-qubit $U(1)$ equivariant gates, defined in Eq.~\eqref{eq:j1j2-u1-circuit}, performs the worst for both values of $\mathcal{J}_2=0.0$ and $\mathcal{J}_2 = 0.44$.
This observation suggests that there is no shortcut in the adiabatic path between $\mathcal{J}_2=0.5$ (which has our initial state $\ket{\psi_0}$ as the ground state) and the considered values of $\mathcal{J}_2\in \{0.0, 0.44\}$ breaking the $\SU(2)/U(1)$ symmetry~\cite{choquette2021quantum,park2024efficient}.

Secondly, the converged energies from the two-qubit vertex gates, Eq.~\eqref{eq:j1j2-two-qubit-vertex}, are subject to large deviations between instances.
This means that the optimization algorithm cannot find good minima in a reasonable time.
We believe that this is due to small initial gradients. 
One can check that the gradient of the cost function at $\vec{\theta}=0$, $\partial_{\theta_{i,j}} \braket{\psi(\{\theta_{i,j}\})|H|\psi(\{\theta_{i,j}\})}|_{\vec{\theta}=0}$, is zero for the second ansatz, while it shows large norm for the first and third ans\"{a}tze.
Since our initialization method (inspired by Ref.~\cite{park2024hamiltonian}) guarantees large initial gradients only when the gradient magnitudes are large enough when $\pmb{\theta}=0$, the second ansatz suffers from the trainability issue.
This fact explains why some of the instances find good minima while others do not.

Thirdly, the converged normalized energies from the ansatz with three-qubit vertex gates are generally closer to the true ground state energy.
Especially when $\mathcal{J}_2=0.44$, the converged energy from the three-qubit vertex gates decreases as the number of parameters increases, whereas that from the two-qubit vertex gates gets flat.
This example shows that using a multi-qubit vertex gate is helpful even for solving a Hamiltonian with two-body interactions.
We expect this because the circuit ansatz with three-qubit vertex gates is more expressive than two-qubit vertex gates when the same number of parameters is provided.

Lastly, the Hamiltonian is one dimensional and is gapped when $\mathcal{J}_2=0.44$ while it is gapless when $\mathcal{J}_2=0.0$.
Therefore, classical algorithms based on the matrix product states (MPSs), such as the density matrix renormalization group (DMRG), are expected to work better for $\mathcal{J}_2=0.44$.
In contrast, our algorithms show the smaller converged normalized energies for $\mathcal{J}_2=0.0$.
All our ans\"{a}tze are efficiently expressed by the MPSs when $p$ is small, but the opposite is not generally true.
This fact suggests that the expressive power of short-depth circuits can be smaller than those of usual tensor network states when symmetry is imposed. 
We still expect that a circuit with a more complex internal structure may have a similar expressivity as the MPS~\cite{haghshenas2022variational}.

\begin{figure}[t]
    \centering
    \includegraphics[width=0.55\linewidth]{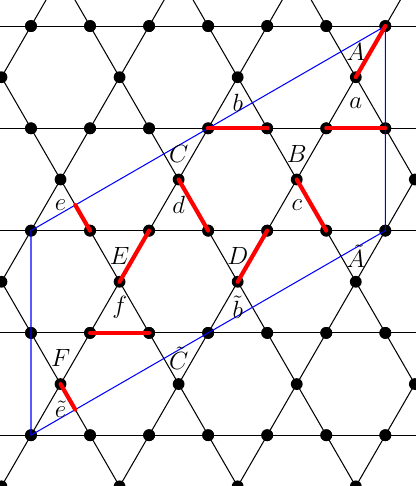}
    \caption{The Kagome lattice. We choose a unit cell with $n=18$ spins enclosed by blue lines.
    Red links indicate the singlets which we use as an initial state. Our variational circuit is constructed by applying three-qubit vertex gates to each triangle ($a$-$f$ and $A$-$F$). See the main text for details.}
    \label{fig:kagome-lattice}
\end{figure}

\subsection{Kagome lattice}
We now extend the previous result to study the model on the Kagome lattice, which is known to be difficult for all known classical algorithms (see, e.g., Refs.~\cite{depenbrock2012nature,suttner2014renormalization,hu2015variational}).
We consider an $n=18$ unit cell from the lattice with periodic boundary conditions.
Our choice of the unit cell is depicted in Fig.~\ref{fig:kagome-lattice}.

Formally, the Hamiltonian of the system is written as
\begin{align}
    H = \sum_{\langle i,j \rangle} \bigl[ X_{i} X_{j} + Y_{i} Y_{j} + Z_{i} Z_{j} \bigr]
\end{align}
where $X_i$, $Y_i$, $Z_i$ are the Pauli operators on site $i$ and the summation is over all nearest neighbors in the lattice.

We construct an ansatz using three-qubit vertex gates as
\begin{align}
    | \psi(\{\theta_{i,j}\}) \rangle = \prod_{i=p}^1 \prod_{j=A}^F V_j(\theta_{i,j}) \prod_{j=a}^f V_j (\theta_{i,j}) | \psi_0 \rangle
\end{align}
where $V_{a,\cdots,f}$ ($V_{A,\cdots,F}$) are the three-qubit vertex gates acting on vertices of each triangle $a$ to $f$ ($A$ to $F$, respectively; see Fig.~\ref{fig:kagome-lattice}).
As each block has $12$ gates, the total number of parameters is $48p$ (recall that each three-qubit vertex gate has four parameters).
We also use a series of singlets as an initial state, where each singlet is indicated by a red link in Fig.~\ref{fig:kagome-lattice}. Formally, we can write
\begin{align*}
    \ket{\psi_0} = \frac{1}{\sqrt{2}^{n/2}} \bigotimes_{\{i,j\} \in S} (\ket{01} - \ket{10})_{ij}
\end{align*}
where $S$ is the set of all links.

\begin{figure}[t]
    \centering
    \includegraphics[width=0.9\linewidth]{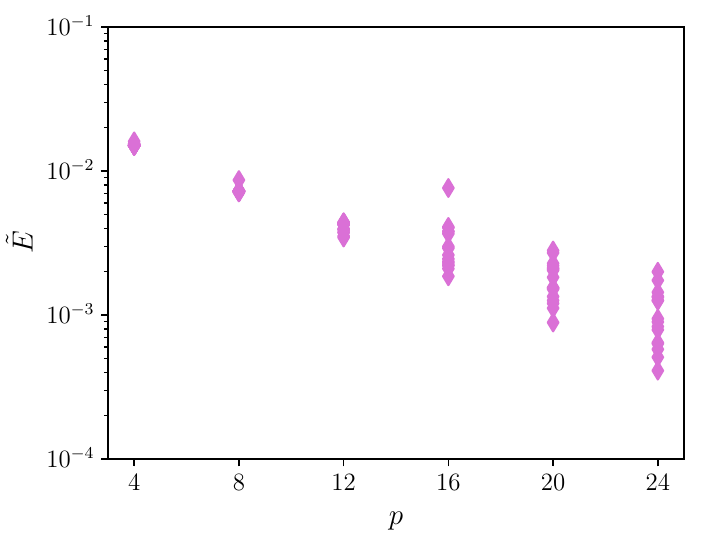}
    \caption{Converged normalized energies as a function of circuit depths for the Heisenberg XXX model on the Kagome lattice. For each value of $p$, $18$ random initial parameters are sampled. A full VQE simulation is performed for each random initial parameter, and the converged energy is shown.}
    \label{fig:kagome18-energy}
\end{figure}

We numerically optimize the parameters of the circuit by minimizing $\langle H \rangle$.
The Adam optimizer is used with the same parameter initialization techniques as in the previous example.
We plot the converged normalized energies as a function of $p$ in Fig.~\ref{fig:kagome18-energy}.
The plot shows that the best-converged energies decrease nearly exponentially with $p$.
The smallest converged normalized energy is $\widetilde{E} \approx 5.7 \times 10^{-4}$ obtained from $p=24$, which is comparable to data obtained in Refs.~\cite{bosse2022probing,kattemolle2022variational} using different ans\"{a}tze.

However, the variance of converged energies increases with $p$.
As the gradient is large for our initial parameters (in contrast to the case we observed for the two-qubit vertex gate with the $J_1-J_2$ model), the deviations between the converged energies suggest that the optimization landscapes can be rugged.
A detailed analysis of the optimization landscapes and the scalability of our algorithm for the Kagome lattice will be presented in future work.

To summarize, we have shown that the three-qubit vertex gate introduced in the previous sections is useful for solving the Heisenberg XXX model on different lattices.
Given the efficacy of our equivariant gates for solving the ground state problem, we also expect that one can construct a QML model using our gates to classify rotationally invariant datasets such as point clouds~\cite{guo2020deep}.
Still, since QML models for those datasets without special pre- and post-processing require a large number of qubits beyond the reach of a classical simulator (which is about $\lesssim 30$ qubits), designing a QML model for a point-cloud dataset will be considered in future work.

Due to the locality constraint, the quantum circuit ans\"{a}tze employed in this section do not necessarily generate the full set of $n$-qubit $\SU(2)$-invariant unitaries~\cite{marvian2022restrictions}. Nevertheless, since we focus on local Hamiltonians, the existence of local gate constructions capable of preparing the ground state is ensured by the adiabatic theorem~\cite{wecker2015progress,hadfield2019quantum}.
Consequently, we expect that the limited expressivity inherent in locally structured ans\"{a}tze does not compromise the circuits’ capacity to accurately represent the ground state of local Hamiltonians.

\section{Connections to PQC, PQC+, and non-classical heuristic algorithms} \label{sec:connections}

Throughout the previous sections, we have introduced an elegant construction method for $\SU(2)$ equivariant quantum circuits based on the Schur transformation.
Those circuits can be naturally seen as spin networks, tensor networks of group-invariant tensors.
We have further developed a theory of the $\SU(2)$ equivariant gates from the Schur-Weyl duality, relating our gates to other known constructions based on the twirling formula and generalized permutations.

As spin networks and quantum circuits for permutations appear in many different contexts in the field of high energy physics and theoretical quantum computations, we discuss various connections to other fields of research as well as possible future directions of study in the following.

The idea of taking spins and coupling them is reminiscent of a computational model already seen in the literature.
This idea is at the heart of what we mentioned above and is called permutational quantum computing (PQC), which is centered around the computational class PQC and the closely related PQC+~\cite{jordan2009permutational,zheng2022super}.
This class of problems is important as it provides strong evidence that the transition from permutations to exponentiated sums of the generators of permutations marks a transition to classically hard sampling tasks.

\paragraph{The PQC model}

In short, PQC is a quantum computing model intimately tied to the structure of a \emph{binary tree} coupling of spins.
The original idea stemmed from the notion that spin networks could form a model of quantum computing~\cite{marzuoli2005computing}. 
To extract a formal computational class from this model, PQC was introduced, which only considers tree-like structures~\cite{jordan2009permutational}.
To achieve this, we take $n$-spins and choose a particular ordering to add the qubits to the already coupled spins (which we can see as a choice of what sequence of spins to apply the $J^2$ operator to).
The possible outcomes of this chosen order of spin recoupling, along with the addition of the possible total angular momentum outcomes, give an alternative basis.

\begin{figure}[t]
    \centering
    \includegraphics[width=0.8\linewidth]{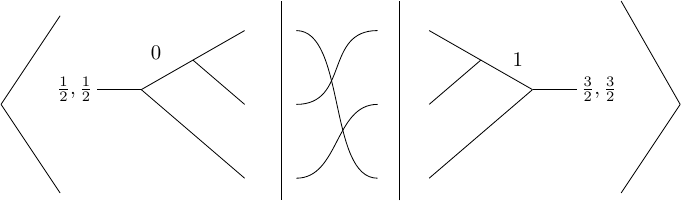}
    \caption{A PQC calculation is an expectation value of a permutation of qubits in the spin-coupling basis.}
    \label{fig:pqc-calc}
\end{figure}

PQC is the computational class of problems described as a permutation circuit set between two coupled spin-basis states.
Given a permutation operator $U_\sigma$ representing the unitary composed of swap gates implementing the permutation $\sigma\in \Sym_n$, PQC is the set of problems written as:
\begin{equation}
    \left\langle{v^\prime}\left|U_\sigma\right| v\right\rangle = \left\langle{b^\prime}\left|S^\dagger U_\sigma S\right| b\right\rangle
\end{equation}
where $b$ is some binary label for the computational basis and $S$ is the Schur gate. The Schur gate is a core component in PQC because PQC states are simply elements of the spin basis.

The Schur gate is the preparation procedure that sends qubit basis states to spin states.
In the PQC literature, these states are often presented by PQC coupling diagrams of the kind seen in Fig.~\ref{fig:trees}.
Practically, a standard PQC calculation is merely the inner product between two Schur gates applied to some computational basis states with some SWAP gates in between them.
It was shown that this model is, in fact, classically simulable in large part due to the particular tree-like structure of binary spin-recoupling and the restrictions this tacitly forces on the Clebsch-Gordan coefficients dictating their coupling~\cite{havlivcek2018quantum}.
An immediate observation we can make, given our above discussion on spin networks, is that PQC diagrams, which we take to be sequentially coupled spin-$1/2$s, are spin networks with their external wires fixed to specific $J_z$ values.
Each PQC basis element is a member of the collection of spin networks of the same tree structure permissible by the recoupling of their spins and a $J_z$ value angular state at the end of the tree.

\paragraph{PQC+}

Despite the initial disappointment that PQC was classically simulable,
it has been generalized to a broader model that is believed to be unlikely to have this property.
The extended model is known as PQC+ where instead of working with a permutation $\sigma \in \Sym_n$, we work with unitaries generated by sums of elements of the permutation algebra $\mathbb{C}[\Sym_n]$: this is composed of elements $f = \sum_i c_i\sigma_i$ with $U_{f}=e^{if} = e^{i\sum_k c_k \sigma_k}$ so, in the end, computations are defined in the following manner:
\begin{equation}
    \left\langle v^\prime\left|U_{f}\right| v\right\rangle.
\end{equation}

\begin{figure}[t]
    \centering
    \includegraphics[width=1\linewidth]{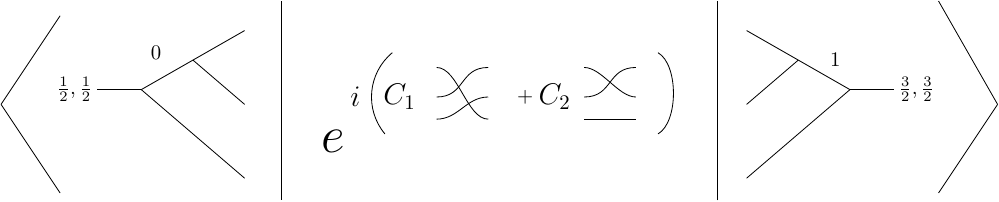}
    \caption{A PQC+ calculation is the exponential of a linear combination of the generators of permutations.
    Previously, in Fig.~\ref{fig:pqc-calc}, the permuted wires stand for the actual permutation, while here they stand for the generators.}
    \label{fig:pqc+calc}
\end{figure}

As was mentioned above, the belief in the resilience of this model to `dequantization' rests on the fact that PQC+ is capable of approximately computing $\Sym_n$ Fourier coefficients in polynomial time; the details can be found in Ref.~\cite{zheng2023speeding}.
The general idea is that, much like in a traditional Fourier transform, to calculate the Fourier coefficient of any element, one must get the component from every element in the original basis, so in the worst case, one must go through as many components as there are basis elements classically.
For an $\Sym_n$ Fourier transform, there is a factorial number of elements~\footnote{This is sloppy, as one runs over the number of irreps which is slightly smaller than permutationally, i.e., factorially, large but remains super-polynomial.}, as such, even an approximate classical polynomial time algorithm to compute the worst case is unlikely. 
This property relates to claims of super-polynomial speedup as permutational complexity grows considerably faster than the exponential. For more details, we direct the reader to Refs.~\cite{zheng2023speeding,zheng2022super}, where one also finds some practical applications in condensed matter calculations in accessing coefficients relevant to the Heisenberg model.

\paragraph{Spin-network circuits as non-classical heuristics}

The central observation in work on PQC+ is that, for a Hamiltonian $H=\sum_i c_i \sigma_i$, we can approximate $\langle u|\exp (-i t H)| v\rangle$ in polynomial time using a quantum circuit.
As the Hamiltonian is in the space $\mathbb{C}\left[\Sym_n\right]$ (the algebra of permutations), we are computing $\mathbb{C}\left[\Sym_n\right]$ Fourier coefficients in polynomial time.
Given that this computation of a Fourier coefficient using the best-known classical algorithm requires one to run over all of $\Sym_n$ that is super-exponential in size, PQC+ allows a super-polyonimal speedup.
This suggests that, in general, elements of the form $\langle u|e^{i\sum_k c_k \sigma_k}| v\rangle$ cannot be efficiently classically computed~\cite{zheng2022super}.
These elements, however, are precisely the form of our parameterized vertex gates---this tells us that the paths through parameter space on which our vertex gates move are classically inaccessible.
This motivates us to introduce the term \emph{non-classical heuristics}---parameterized ans\"{a}tze that are defined as moving through spaces that cannot be accessed classically in polynomial time. 
For example, if we consider a random circuit sampling, but limited to $\SU(2)$-equivariant circuit, we still expect that this problem would be difficult (as for quantum Fourier sampling~\cite{fefferman2015power}).

However, we should note that this idea does not tell us if randomly moving through this space is useful; the space may still be barren~\cite{mcclean2018barren}.
The question, then, is that whether there is practical problems, such as $\SU(2)$ equivariant optimization problems, for which we can design quantum circuit heuristics, such as spin-network circuits, that is not efficiently classically simulable but still trainable.

In terms of the approaches to machine learning presented in PQC+ to date and our spin-network circuits, it should be noted that there is a technical distinction between the methods used. The PQC+ focuses on tuning the coefficients $c_i$ of the exponent $\sum_i c_i \sigma_i\in \mathbb{C}\left[\Sym_n\right]$. In our spin-network circuits, we parameterize the distinct $\SU(2)$ spin spaces and mix spin irreps of the same $J$-value in the Schur-Weyl decomposition via unitaries (see Corollary~\ref{col:unitary_commutator}). Though both exist in the same space, the way in which one moves through that parameter space is very different.

\section{Conclusion} \label{sec:conclusion}

In this paper, we have put forward theoretically motivated ans\"{a}tze based on spin networks, a form of $\SU(2)$ equivariant tensor networks.
This offers a way to design $\SU(2)$ equivariant variational quantum algorithms, which are natural for rotationally invariant quantum systems, based on the Schur map induced by a spin-coupling diagram.
Furthermore, we have shown that our approach yields the same parameter spaces as the twirling formula, but in a direct manner that avoids the twirling computation for many-qubit gates, which is highly non-trivial.

For the two and three-qubit gate cases, we have further justified our approach by classically simulating the variational quantum eigensolver (VQE) for finding the ground state of the $\SU(2)$ symmetric Heisenberg (XXX) models on the one-dimensional triangular (for $20$ qubits) and on the Kagome (for $18$ qubit unit cell) lattices. 
The results show that our three-qubit $\SU(2)$-equivariant gate can be advantageous over the previously studied two-qubit equivariant gates. 
We note that the scalability of the algorithm (especially for the Kagome lattice) for larger systems still needs to be clarified in future work, since the VQEs' performance depends on multiple factors such as the initial parameter conditions and the optimization landscapes.

Connecting to the broader literature, we also have shown that $\SU(2)$ equivariant gates are identical to the generalized permutations discussed in the context of PQC+ \cite{zheng2023speeding}.
The connection to PQC+ is also used to argue how our ans\"{a}tze move through a parameter space that a classical algorithm finds difficult to access.
The original observation in Ref.~\cite{zheng2023speeding} showed that the expectation value of generalized permutations in the spin-basis calculates $\Sym_n$ Fourier coefficients in polynomial time (a possible super-polynomial speedup), and our work now extends this to $\SU(2)$ equivariant gates.
This leads to our introduction of the term \emph{non-classical heuristics} for quantum variational techniques, which can be argued to access regions of the parameter space that are classically intractable.

Future research in this direction may extend this notion to rigorous complexity arguments by finding a task with $\SU(2)$ symmetry that is solvable by $\SU(2)$ equivariant circuits where no known efficient classical algorithm exists.
For example, Ref.~\cite{liu2021rigorous} has proven quantum advantage in an ML task by designing a dataset whose classification task is convertible to the discrete logarithm problem, which is efficiently solvable by a QML algorithm, yet an efficient classical algorithm is deemed impossible (unless the discrete logarithm problem is in \textsf{BPP}).
Similarly, we expect it is possible to design an ML task related to the Fourier transformation over $\Sym_n$, also establishing rigorous quantum advantage arguments in this domain.

We also expect that the Schur transformation method introduced here---which changes to the basis block diagonalizing $\SU(2)$ group action---can be used to simulate the dynamics of $\SU(2)$-symmetric Hamiltonians.
The quantum circuit ans\"{a}tze used in this work [e.g., Eq.~\eqref{eq:j1j2-two-qubit-vertex}] are already suitable for variational time evolution~\cite{benedetti2021hardware,lin2021real}, and inherently preserve the $\SU(2)$-symmetry, making them promising candidates for near-term quantum devices.
Moreover, we believe our approach can serve as a foundation for developing fault-tolerant algorithms for simulating the time evolution of a quantum system.
Existing quantum algorithms for time evolution, such as the Suzuki-Trotter decomposition, often break symmetry~\cite{tran2021faster}.
Since time evolution generated by an $\SU(2)$-symmetric Hamiltonian must be an equivariant form (see Corollary~\ref{col:unitary_commutator}) in the Schur basis, adopting this basis may enable quantum-circuit implementations that maintain the symmetry throughout the evolution.
The development of such an algorithm can be impactful not only for the simulation of many-body systems but also for applications in high-energy physics, where the symmetric Hamiltonians are prevalent~\cite{bauer2023quantum}.

\section*{Acknowledgements}
RDPE would like to acknowledge useful conversation and comments from Sergii Strelchuk, Deepak Vaid, and Pierre Martin-Dussaud.
CYP thanks Seongwook Shin for his helpful comments.
All authors thank the Xanadu QML and software research teams, with special thanks to Maria Schuld, David Wierichs, Joseph Bowles, and David Wakeham.
This research was supported (in part) by the Yonsei University Research Fund of 2024-22-0501.
This research was supported (in part) by the BK21FOUR (Fostering Outstanding Universities for Research), funded by the Ministry of Education (MOE) of Korea and the National Research Foundation (NRF) of Korea; and the NRF grants funded by the Korean government (MSIT) (No. RS-2025-16066935, RS-2025-02316431, and  RS-2025-18362970).
This research used resources from the National Energy Research Scientific Computing Center, which is supported by the Office of Science of the U.S. Department of Energy under Contract No. DE-AC02-05CH11231 using NERSC award
NERSC DDR-ERCAP0029552.

\appendix
\numberwithin{equation}{section}

\section{Formal introduction to spin networks}\label{sect:q-geom}

Despite having a modest presentation, the gate architectures seen in Sec.~\ref{sec:spin-network-circuits} cannot be understood beyond a superficial depth without grasping the motivating concept of the spin network more deeply. The spin network can be seen as a type of tensor network where the vertices are invariant under $\SU(2)$ actions, and the contraction edges are indexed by irreps of $\SU(2)$. This relates to a particular representation of equivariant linear maps as \emph{harmonic tensor networks} over $\SU(2)$~\cite{martin2019primer}.
Here, `harmonic` is used in the sense of harmonic analysis and decomposition of functions over representations, and we mean a tensor network where the tensors involved are all equivariant with respect to the given group. 
We will give the classical presentation of a spin network as a labeled graph to allow the interested reader to follow the spin network literature more easily.

\paragraph{Labeled Directed Graphs.}
A spin network is a particular form of a labeled directed graph
A directed graph $\Gamma$ is an ordered pair $\Gamma=(\mathcal{N}, \mathcal{L})$, where $\mathcal{N}=\left\{n_1, \ldots, n_N\right\}$ is a finite set of $N$ nodes, and $\mathcal{L}=\left\{l_1, \ldots, l_{\mathrm{L}}\right\}$ a finite set of $L$ edges (traditionally referred to as links in the loop quantum gravity literature), endowed with a target map $\mathrm{t}: \mathcal{L} \rightarrow \mathcal{N}$ and a source map $\mathrm{s}: \mathcal{L} \rightarrow \mathcal{N}$, assigning each edge to its end and start points, respectively.
We denote $\mathcal{S}(n)$ (respectively $\mathcal{T}(n)$) the set of edges for which the node $n$ is the source (respectively the target). The valency of a node $n$ is the number of edges incident on $n$, i.e., $|\mathcal{S}(n)|+|\mathcal{T}(n)|$. A graph is said to be $p$-valent if the valency of each node is $p$.

\paragraph{Intertwiners.}
Before defining spin networks by restricting ourselves to labeled directed graphs of a certain type, it will be profitable to define the concept of intertwiners.
Let us say that we have two vector spaces $V$ and $W$ on which representations $U_V, U_W:G\rightarrow \mathrm{GL}(V)$ of a group made up of elements $g\in G$.
Then, an intertwiner is a linear map $T: V \rightarrow W$ which satisfies:
\begin{equation}
T(U_V(g)\circ v) = U_W(g) \circ  T (v)
\end{equation}
where $v\in V$. This is alternatively characterized by the commuting diagram:
\begin{equation}
\tikzcdset{diagrams={nodes={inner sep=5pt}}}
\begin{tikzcd}[row sep=large, column sep = large]
	V \arrow{r}{T} \arrow[swap]{d}{U_V(g)} & W \arrow{d}{U_W(g)} \\
	V \arrow{r}{T}& W.
\end{tikzcd}
\end{equation}

This shows us that an intertwiner is an equivariant map. This is also referred to as a covariant map, depending on the literature.

The space of intertwiners denoted $\operatorname{Hom}_{G}(V, W)$, is a subspace of the vector space of linear maps $\operatorname{Hom}(V, W)$ from $V$ to $W$. Given a space of equivariant maps under the group $G$ we can make the following useful identification between the \emph{equivariant} maps with an isomorphic space of \emph{invariant} states
\begin{equation}
\operatorname{Hom}_{G}(V, W) \cong \operatorname{Inv}_{G}\left(V \otimes W^*\right),   
\end{equation}
where $W^*$ is the dual space of $W$ made up of maps from $W$ to the complex numbers. Here, we define an invariant space as
\begin{equation}
\operatorname{Inv}_{G}(E) \stackrel{\text { def }}{=}\{\psi \in E \mid \forall g \in G, g \cdot \psi=\psi\}.
\end{equation}
We can see by the construction from $G$ equivariant maps that the states in $E$ when acted on by $G$ via the representation $U_V\otimes U_W^{\dagger}$ must be such that for any $v\otimes w^\dagger \in V \otimes W^*$
\begin{equation}
    (U_V \otimes U_W^{\dagger}) v\otimes w^\dagger =  (\Id \otimes U_W U_W^{\dagger}) v\otimes w^\dagger = v\otimes w^\dagger
\end{equation}
which is the source of their invariance.

Let us consider again the directed graph $\Gamma$.
We denote $\Lambda_\Gamma$ by the set of labelings $j$ that assign to any edge $l \in \mathcal{L}$ an $\SU(2)$ irreducible representation characterized by the spin number $j_l \in \mathbb{N} / 2$.
Given a labelling $j \in \Lambda_{\Gamma}$, we write
\begin{equation}
\operatorname{Inv}(n, j) \stackrel{\text { def }}{=} \operatorname{Inv}_{\SU(2)}\left(\bigotimes_{l \in \mathcal{S}(n)} V_{j_l} \otimes \bigotimes_{l \in \mathcal{T}(n)} V_{j_l}^*\right),
\end{equation}
where the $V_{j_l}$ are the spaces of the irreps $j_l$ associated with the edges. Using the concept of invariant subspace, we can now define a spin network.

\paragraph{Spin networks.} A \emph{spin network} is a triple $\Sigma=(\Gamma, j, \imath)$, where $\Gamma$ is a directed graph, with a labelling $j \in \Lambda_{\Gamma}$ on its edges, and a map $\iota$ that assigns to every $n \in \mathcal{N}$ an intertwiner $\left|l_n\right\rangle \in \operatorname{Inv}(n, j)$.

\paragraph{Clebsch-Gordan coefficients and the vertex basis}
Having described the spin network abstractly, it can be practical to choose a specific basis in order to look at how the vertices are represented as matrices.
The smallest possible non-trivial intertwiner is three-valent, and we shall see that we can construct all larger valences from these.
For the three-valent intertwiner, the space is $\mathrm{Inv}_{\SU(2)}\left(\mathfrak{J}_{j_{1}} \otimes \mathfrak{J}_{j_{2}} \otimes \mathfrak{J}_{j_{3}}\right)$ (we have dropped the reference to the last space is dual, this is common in the literature as they are isomorphic) and it can be given a basis by sequentially coupling the first two spins and then contracting the result with the third. Firstly, we need to map the tensor product of the first two spins $\mathfrak{J}_{j_{1}} \otimes \mathfrak{J}_{j_{2}}$ to the direct sum basis $\mathfrak{J}_{j_{1}} \oplus \mathfrak{J}_{j_{2}}$ as in 
\begin{align}
\mathfrak{J}_{j_1}\otimes \mathfrak{J}_{j_2} \simeq \bigoplus_{k=|j_1-j_2|}^{j_1+j_2} \mathfrak{J}_k \label{eq:decomp}
\end{align}
Here the equivalence is given by the intertwiner map:
\begin{align}
    \iota \left\{
    \begin{array}{l} \mathfrak{J}_{j_1}\otimes \mathfrak{J}_{j_2}\rightarrow \bigoplus_{k=\left|j_1-j_2\right|} \mathfrak{J}_k \\
          |j_1,j_2;m_1 m_2 \rangle \rightarrow |km \rangle
    \end{array} \right.
\end{align}

Written in this form, we can see that the intertwiner map is a change of basis to block diagonalising the representation, and each block is an irreducible representation. 
This is just the Schur map when we have qubits, i.e., spin-$1/2$s as the first two spaces.
The matrix coefficients of the map $\iota$ are given by the Clebsch-Gordan coefficients~\footnote{In the spin network literature, we often see that vertices are described via Wigner symbols instead of Clebsch-Gordan coefficients as seen here. The Wigner symbols are an equivalent way to decompose three vector spaces as is done here, which is more symmetric. Since we are looking to derive computations with well-defined input and output, it is simpler to use this basis instead. See Ref.~\cite{east2021spin} for more details.}
\begin{align}
C^{jm}_{j_1\:m_1\:j_2\:m_2}:=\langle j_1m_1;j_2m_2|jm\rangle \label{eq:cg}
\end{align}

Clebsch-Gordan coefficients are usually first encountered by physicists during undergraduate courses in atomic physics. They are typically presented as the obscure coefficients that dictate how different (atomic) spin states $\ket{j_1,m_1}$ and $\ket{j_2,m_2}$ combine together to form a combined $\ket{j,m}$ state as seen in the equation:
\begin{equation}
    |j m\rangle=\sum_{m_1=-j_1}^{j_1} \sum_{m_2=-j_2}^{j_2}c^{jm}_{j_1,j_2,m_1, m_2}\left|j_1 m_1 j_2 m_2\right\rangle,
\end{equation}
where the coefficients are taken to be non-zero only when the Clebsch-Gordan conditions hold:
\begin{equation}\label{clebsch-base}
    \begin{aligned}j_1+j_2+j\in\mathbb{N}\\ \lvert j_1-j_2\rvert\leqslant j\leqslant j_1+j_2.\end{aligned}
\end{equation}

Notably, the space $\mathrm{Inv}_{\SU(2)}\left(\mathfrak{J}_{j_{1}} \otimes \mathfrak{J}_{j_{2}} \otimes \mathfrak{J}_{j_{3}}\right)$ is one dimensional, meaning there is only one intertwiner up to a scalar. This makes sense because, in the three-valent case, the choice of two spins completely fixes the third~\cite{martin2019primer}.

For higher valence networks, we can build a similar basis by reapplying the decomposition procedure seen in Eq.~\eqref{eq:decomp} until all the tensor products are replaced by direct sums. For example, in the case of four-valent spin networks, we reapply Eq.~\eqref{eq:decomp} to three-valent product spaces tensored with the third spin
\begin{equation}
    \left(\bigoplus\limits_{{k}=|j_1-j_2|}^{j_1+j_2} \mathfrak{J}_k \right) \otimes \mathfrak{J}_{j_{3}}=\bigoplus\limits_{j_{12}=|j_{1}-j_{2}|}^{j_{1}+j_{2}}\bigoplus_{k=|j_{12}-j_{3}|}^{j_{12}+j_{3}} \mathfrak{J}_{{k}}.
\end{equation}
This, in terms of states and Clebsch-Gordan coefficients, leads to the following:

\begin{align}
    &\ket{(j_1j_2)j_3;j_{12}kn} \nonumber \\
    &= \sum_{m_1,m_2,m_3,m_{12}}C_{j_1m_1j_2m_2}^{j_{12}m_{12}}C_{ j_{12}m_{12}
  j_3m_3}^{k n}\bigotimes\limits_{i=1}^3|j_i,m_i\rangle. \label{cg:decomp} 
\end{align}
It is important to note that there is freedom in ordering the breakdown of a tensor product of three spaces into direct sums.
Here, we take the first two spins, consider the resultant direct sum, and then take the tensor product with the third space.
This order could be reversed; one could take the second and third or the first and third.
These separate decompositions amount to different basis choices that play a role in the structure of permutational quantum computing discussed above (see Sec.~\ref{sec:connections}).
As a side note, the quantum gravity community is mostly interested in three- and four-valent spin networks due to a relationship with 2D and 3D space models of gravity~\cite{rovelli2015covariant,martin2019primer}.
In addition, there is a direct relationship with the present quantum computing literature and three-valent intertwiners due to the work on PQC.

\section{The representation theory of the symmetric group}\label{app:rep_sym}

In this Appendix, we briefly introduce the irreducible representation of the symmetric group $\Sym_n$.

Consider a partition of a positive integer $n$ to a monotonically decreasing sequence of positive integers, $\lambda=(\lambda_1,\lambda_2,\cdots)$ that sum to $n$. We can associate these integers with cycle shapes of $\Sym_n$. For example, given ten elements, we can associate the partition $\lambda = (4,2,2,2)$ with a permutation decomposable into one four-cycle and three two-cycles.

A Young diagram is a diagrammatic depiction of the cycle shapes of $\Sym_n$. Typically, the largest cycle goes at the top, and for every element in the cycle, we add a box, as seen here:
\begin{align}
    \ytableausetup{centertableaux,boxsize=1em}
    \ydiagram{4,2,2,2} 
\end{align}

A \emph{Standard filling} of a Young diagram is a bijective map of the numbers from $1$ to $n$, where n is the number of boxes such that the entries increase along the rows and down the columns. The standard filled Young diagram is called a \emph{Young tableau}

\begin{align}
    \ytableausetup{centertableaux}
    \begin{ytableau}
    1 & 3 & 4 & 7\\
    2 & 8 \\
    5 & 9 \\
    6 & 10\\
    \end{ytableau}
\end{align}

We can apply an element of the symmetric group to the tableau by simply applying the permutation $\alpha\in \Sym_n$ to the numbers in the tableau.

Let us define the equivalence class $R(T)$ of permutations that only move elements about \emph{within their rows}. In this way, we define the row stabilizers, simply the product subgroups $\bigotimes_{p\in \lambda} S_p$. In our earlier example, it would be the space $S_4\otimes S_2 \otimes S_2 \otimes S_2$. Analogously, we can also describe the column stabilizers $C(T)$.

To describe the irreps of $\Sym_n$, we will need the Young polytabloid:
\begin{equation}
    e_T = \{T\}=\sum_{\alpha\in C(T)}\operatorname{sgn}(\alpha)\alpha\triangleright T
\end{equation}
where $\operatorname{sgn}(\alpha)$ is the parity function giving $1$ for an even permutation or $-1$ for an odd one. We note that $\alpha\triangleright T$ is not necessarily a Young tableau due to its non-standard filling.

For example, given the tableau
\begin{align}
    \ytableausetup{centertableaux}
    \begin{ytableau}
    1 & 2\\
    3 \\
    \end{ytableau}
\end{align}
the polytabloid is given by
\begin{align}
    \left\{\begin{ytableau}
    1 & 2\\
    3 \\
    \end{ytableau}\right\} &= \mathrm{sgn}(Id)\ \begin{ytableau}
    1 & 2\\
    3 \\
    \end{ytableau} + \mathrm{sgn}((1,3))\ \begin{ytableau}
    3 & 2\\
    1 \\
    \end{ytableau} \\
    &= \begin{ytableau}
    1 & 2\\
    3 \\
    \end{ytableau} - \begin{ytableau}
    3 & 2\\
    1 \\
    \end{ytableau}
\end{align}

A Specht module is a module spanned by polytabloids $e_T$ where $T$ is the index corresponding to all tableaux of shape $\lambda$. That is to say.
\begin{align}
    Sp^{(\lambda)} &= \{c_{1}e_{T_{1}}+c_{2}e_{T_{2}}+c_{3}e_{T_{3}}+\cdots{|}c_1,c_2\ldots\in\mathbb{C} \nonumber \\
    & \qquad \text{ and } T_1,T_2\ldots \text{ are tableaux of shape } \lambda \}.
\end{align}
It can be shown that the Specht modules are the irreps of $\Sym_n$~\cite{mcnamara2013irreducible}.

In the context of the above work, let $n=3$, and restrict to the Young diagrams with at most two rows which correspond to the multiplicity of elements of $\SU(2)$ by Schur-Weyl duality. These are

\begin{align}
    \ytableausetup{centertableaux,boxsize=1em}
        \ydiagram{3} 
        \qquad \text{ and } \qquad  \ydiagram{2,1}\,.
\end{align}
The irreducible representations of $S_3$ associated with the first diagram are dimension $1$, and the second diagram is dimension $2$. More precisely, the Specht module for the first diagram is generated by a single vector:

\begin{equation}
    \left\{\begin{ytableau}
    1 & 2 & 3
    \end{ytableau}\right\}= \begin{ytableau}
    1 & 2 & 3
    \end{ytableau}.
\end{equation}

The Specht module for the second diagram is generated by two vectors which correspond to the two possible tableau
\begin{equation}
    \left\{\begin{ytableau}
    1 & 3\\
    2 \\
    \end{ytableau}\right\}= \begin{ytableau}
    1 & 3\\
    2 \\
    \end{ytableau} + \begin{ytableau}
    2 & 3\\
    1 \\
    \end{ytableau}
\end{equation}
and
\begin{equation}
    \left\{\begin{ytableau}
    1 & 2\\
    3 \\
    \end{ytableau}\right\}= \begin{ytableau}
    1 & 2\\
    3 \\
    \end{ytableau} - \begin{ytableau}
    3 & 2\\
    1 \\
    \end{ytableau} .
\end{equation}

Referring back to the Schur-Weyl decomposition where the irreps of $\Sym_n$ give the multiplicities of the $\SU(2)$ irreps, we observe:
\begin{align}
        (\mathbb{C}^2)^{\otimes 3} \simeq \mathfrak{J}_{3/2} \oplus 2 \mathfrak{J}_{1/2},
\end{align}
as the three-row element corresponds to the fully symmetric subspace of the three-qubit components, i.e., spin-$3/2$ and the mixed representation corresponds to the spin-$1/2$. For more details, see Refs.~\cite{sagan2013symmetric,fulton2013representation}.

\section{Further notes on the Schur gate}

\paragraph{Equivariance of the Schur gate}

Focusing on the Schur matrix in Eq.~\eqref{eq:schur-map}, a natural question is: How are the representations of the group acting on the input affected by the Schur map?
As discussed above, the input space has the tensor product representation, and the output has the spin representation, which functions differently.
A useful shorthand to express the idea of a group element $g$ acting on some space $H$ without worrying about how exactly it is represented is to write $g\triangleright H$.
With this in mind let us consider the action of $\SU(2)$ for the two qubit case, for an arbitrary element $g\in \SU(2)$, the input space of the Schur map will transform as $g\triangleright (\mathbb{C}^2\otimes \mathbb{C}^2)= (g\triangleright \mathbb{C}^2)\otimes (g\triangleright \mathbb{C}^2) = U_g \otimes U_g $, where $U_g$ is the qubit representation of the element $g$.
The output space, however, will transform differently as we are viewing the space as composed of spin components, $g\triangleright (\mathfrak{J}_0\oplus \mathfrak{J}_1)= (g\triangleright \mathfrak{J}_0)\oplus( g\triangleright \mathfrak{J}_1)=\mathfrak{J}_0\oplus( g\triangleright \mathfrak{J}_1)$, which implies that $g$ acts as $\Id \oplus \pi^1(g)$, where the action on the single element spin-$0$ subspace is trivial and $\pi^1(g)$ is the spin-$1$ representation of the element $g$.
We can use the Schur map itself as a mapping between the tensor product basis and the spin space to create a representation on the direct product, i.e., $U(g)^{\otimes k} = S^\dagger \pi(g)S$, which we can see as mapping our tensor space to a spin space, performing the group action there, and then sending it back.
Let us now see that our Schur map $S$ is equivariant under the action of $g$, which if from a direct and short calculation:
\begin{equation}
  S_2 (\Id \oplus \pi^1(g)) = S_2( S_2^\dagger U(g)^{\otimes k} S_2)=  U_g^{\otimes k} S_2.
\end{equation}
The group action has moved from the right-hand side of the Schur gate to the left, and so they commute, which is the definition of equivariance.
This calculation, though short and can be somewhat deceptive, it is imperative that we remember that the action of the group should be represented differently before and after the Schur gate. The effect of placing the group action between the Schur gates was to transform it into the appropriate action on the spin space.

A similar discussion applies to the three-qubit space. Recalling that $\mathbb{C}^2\otimes \mathbb{C}^2\otimes\mathbb{C}^2\simeq \mathfrak{J}_{3/2}\oplus \mathfrak{J}_{1/2}\oplus \mathfrak{J}_{1/2}$ we would then say that $g\in G$ acts as $g\triangleright (\mathfrak{J}_{3/2}\oplus \mathfrak{J}_{1/2}\oplus \mathfrak{J}_{1/2}) = (g\triangleright \mathfrak{J}_{3/2})\oplus (g\triangleright \mathfrak{J}_{1/2})  \oplus (g\triangleright \mathfrak{J}_{1/2})$ and in the end we have that we can use the Schur gate to map us between representations acting on these spaces:
\begin{equation}
    S_3 (\pi^{\frac{1}{2}}(g) \oplus \pi^{\frac{1}{2}} \oplus \pi^{\frac{3}{2}}(g)) =S_3(S_3^\dagger U(g)^{\otimes k}S_3)= U(g)^{\otimes k}S_3.
\end{equation}
Indeed, this structure will hold in general.

\paragraph{The Schur gate and PQC recoupling diagrams}
    
As elements of the spin-basis, the PQC diagrams exactly correspond to the elements of the Schur basis. When specific $J_z$ values are fixed on all the external wires, one can use the PQC diagrams to index the Schur matrix.

For example, let us recall the $S_2$ matrix which is given by
\begin{align}
S_2 &= 
\begin{pmatrix}
1 & 0 & 0 & 0\\
0 & \frac{1}{\sqrt{2}} & \frac{1}{\sqrt{2}} & 0\\
0 & 0 & 0 & 1\\
0 & \frac{1}{\sqrt{2}} & -\frac{1}{\sqrt{2}} & 0
\end{pmatrix} \\
&= \begin{pmatrix}
c^{1,1}_{\frac{1}{2},\frac{1}{2};\frac{1}{2},\frac{1}{2}} & c^{1,1}_{\frac{1}{2},\frac{1}{2}; \frac{1}{2},-\frac{1}{2}}& c^{1,1}_{\frac{1}{2},-\frac{1}{2}; \frac{1}{2},\frac{1}{2}}& c^{1,1}_{\frac{1}{2},-\frac{1}{2};\frac{1}{2},-\frac{1}{2}}\\
c^{1,0}_{\frac{1}{2},\frac{1}{2};\frac{1}{2},\frac{1}{2}} & c^{1,0}_{\frac{1}{2},\frac{1}{2}; \frac{1}{2},-\frac{1}{2}}& c^{1,0}_{\frac{1}{2},-\frac{1}{2};\frac{1}{2},\frac{1}{2}}& c^{1,0}_{\frac{1}{2},-\frac{1}{2};\frac{1}{2},-\frac{1}{2}}\\
c^{1,-1}_{\frac{1}{2},\frac{1}{2};\frac{1}{2},\frac{1}{2}} & c^{1,-1}_{\frac{1}{2},\frac{1}{2};\frac{1}{2},-\frac{1}{2}}& c^{1,-1}_{\frac{1}{2},-\frac{1}{2};\frac{1}{2},\frac{1}{2}}& c^{1,-1}_{\frac{1}{2},-\frac{1}{2};\frac{1}{2},-\frac{1}{2}}\\
c^{0,0}_{\frac{1}{2},\frac{1}{2};\frac{1}{2},\frac{1}{2}} & c^{0,0}_{\frac{1}{2},\frac{1}{2};\frac{1}{2},-\frac{1}{2}}& c^{0,0}_{\frac{1}{2},-\frac{1}{2};\frac{1}{2},\frac{1}{2}}& c^{0,0}_{\frac{1}{2},-\frac{1}{2};\frac{1}{2},-\frac{1}{2}}
\end{pmatrix}.
\end{align}
Each element of the matrix can be seen as a PQC diagram 
\begin{equation}
c^{J,M}_{\frac{1}{2},m_1;\frac{1}{2},m_2} = \begin{tikzpicture}[baseline,none/.style={}]
    \node[none] (0) at (-2, 1) {$\frac{1}{2},m_1$};
    \node[none] (1) at (-2, -1) {$\frac{1}{2},m_2$};
    \node[none] (2) at (-0.5, 0) {};
    \node[none] (3) at (2.5, 0) {$J,M$};

    \draw (0.east) to (2.center);
    \draw (1.east) to (2.center);
    \draw (2.center) edge ["$J$"] (3);
  \end{tikzpicture}
\end{equation}

The connection becomes clearer in the three-qubit case, showing how the entries of the matrices are the combinations of Clebsch-Gordan coefficients that correspond to particular coupling structures:
\begin{equation}
\begin{aligned}
    S_{3} &= (c^{j_4,m_4}_{j_1,m_1;j_2,m_2}c^{J,M}_{j_4,m_4;j_3,m_3}) \\
    &= \begin{pmatrix}
    1 & 0 & 0 & 0 & 0 & 0 & 0 & 0\\
      0 & \frac{1}{\sqrt{3}} & \frac{1}{\sqrt{3}} & 0 & \frac{1}{\sqrt{3}}& 0 & 0 & 0 \\
       0 & 0 & 0 & \frac{1}{\sqrt{3}} & 0 & \frac{1}{\sqrt{3}} & \frac{1}{\sqrt{3}}  & 0 \\
       0 & 0 & 0 & 0 & 0 & 0 & 0 & 1\\
       0 & \sqrt{\frac{2}{3}} & -\frac{1}{\sqrt{6}} & 0 & -\frac{1}{\sqrt{6}} & 0 & 0 & 0\\
       0 & 0 & 0 & -\frac{1}{\sqrt{6}} & 0 & -\frac{1}{\sqrt{6}} & \sqrt{\frac{2}{3}} & 0 \\
       0 & 0 & \frac{1}{\sqrt{2}} & 0 & \frac{1}{\sqrt{2}} & 0 &0 & 0\\
       0 & 0 & 0 & -\frac{1}{\sqrt{2}} & 0 & \frac{1}{\sqrt{2}} & 0 & 0
\end{pmatrix} \nonumber\Leftrightarrow 
\end{aligned}
\end{equation}
\begin{equation}
\begin{aligned}
&\begin{tikzpicture}[baseline,none/.style={}]
    \node [style=none] (0) at (-2, 0.75) {$\{\frac{1}{2},-\frac{1}{2}\}$};
    \node [style=none] (1) at (-2, -0.75) {$\{\frac{1}{2},-\frac{1}{2}\}$};
    \node [style=none] (2) at (-2, -2) {$\{\frac{1}{2},-\frac{1}{2}\}$};
    \node [style=none] (3) at (-0.5,0) {};
    \node [style=none] (4) at (1,0) {};
    \node [anchor=west,style=none] (5) at (3.0,0) {$\{\frac{3}{2},\frac{1}{2},-\frac{1}{2},-\frac{3}{2}\}$};

    \draw (0.east) to (3.center);
    \draw (1.east) to (3.center);
    \draw (3.center) edge ["{\{0,1\}}"] (4.center);
    \draw (2.east) to (4.center);
    \draw (4.center) edge ["{$\{\frac{3}{2},\frac{1}{2}g1,\frac{1}{2}g0\}$}"] (5.west);
\end{tikzpicture}
\label{eq:schur3}
\end{aligned}
\end{equation}
The terms $\frac{1}{2}g1$ and $\frac{1}{2}g0$ serve to separate the two ways one can couple to a total angular momentum of $\frac{1}{2}$ on the last edge. Specifically, $\frac{1}{2}g1$ indicates the case when the initial coupling resulted in a total angular momentum of $1$, and $\frac{1}{2}g0$ is for when it resulted in $0$. These have to be distinguished as they correspond to the multiplicities of spin-$\frac{1}{2}$ and so do actually index different elements in the matrix. Here, we merely state the $J_z$ values at the sides of the wires on the RHS, and we assume the $J_z$ values range only where permissible.

\bibliography{bib.bib}

\end{document}